\documentclass{article}
\usepackage{amsmath}
\usepackage{amsfonts}
\usepackage{graphicx}
\usepackage{xcolor}
\usepackage[title]{appendix}

\usepackage{cite}

\begin{document}

\nocite{*}

\section*{Magnon Hall effect in antiferromagnetic lattices}

\noindent P. G. de Oliveira$^a$* \newline
\noindent E-mail address: pgopedro@ufmg.br  \newline

\noindent and \newline

\noindent A. S. T. Pires$^a$ \newline
\noindent E-mail address: antpires@fisica.ufmg.br
\vspace{\baselineskip}

\noindent $^a$ Departamento de Física, Universidade Federal de Minas Gerais, Belo Horizonte, MG, CP702, 30123-970, Brazil.
\noindent \newline * Corresponding author.

\subsection*{Abstract}

Topology applied to condensed matter is an important area of research and technology, and topological magnetic excitations have recently become an active field of study. This paper presents a general discussion of magnon Hall transport in two-dimensional antiferromagnets. \textcolor{black}{Although the Chern number is zero for a collinear antiferromagnet, we offer a general discussion that can be used in the more general case.} First, we study the Union Jack lattice, where an effective time-reversal symmetry is broken, making the system display the magnon Hall effect. Then, we investigate the brick-wall lattice where such symmetry is present. Consequently, we have a phenomenon similar to the quantum spin Hall effect in electronic systems. Both lattices have not yet been studied from the topological point of view. The coexistence of opposite spin polarization in an antiferromagnet resembles the electron spin in various transport phenomena. \textcolor{black}{We study magnon transport} in the lattices mentioned above with Dzyaloshinskii-Moriya interaction and easy-axis single-ion anisotropy. We calculate the Berry curvature from the eigenvalues of the Hamiltonian. From that, we plot the spin Hall and thermal Hall conductivities, as well as the spin Nernst coefficient, as functions of the temperature. In the Union Jack lattice, we treat the effect of anharmonic interactions using a mean-field spin wave theory where the Hamiltonian becomes implicitly temperature-dependent. We determine self-consistently the renormalized dispersion and the staggered magnetization as a function of temperature. Our calculations can be applied to other antiferromagnetic lattices. 

\vspace{\baselineskip}

\noindent Keywords: topological magnons, spin Nernst effect, thermal Hall effect, magnon transport

\newpage
\normalsize

\section{Introduction}

Over the past decades, topology has played an essential role in condensed matter physics, and one of the most studied phenomena is the quantum Hall effect (QHE). Suppose we have electrons localized in the $xy$-plane subjected to a magnetic field pointing in the up $z$-direction. We consider also a current $I_x$ along the $x$-direction, associated with a current density $j_x$ and an electric field $E_x$. The transversal magnetic field deflects the electrons sideways in the negative $y$-direction. If the sample is infinite in the $y$-direction, the electrons will move with some angle relative to the $x$-axis. But if the sample is finite in the $y$-direction, the deflected electrons will accumulate on the edge of the sample, and will produce a transversal electric field $E_y$. In a sample with a strip geometry, the QHE makes electrons at the top edge flow to the right while electrons at the bottom move to the left. We have $J_\alpha = \sigma_{\alpha \beta}E_\beta$. The time-reversal (TR) operation inverts $J_\alpha$ but not $E_\beta$. Therefore, $\sigma_{\alpha \beta}$ must break time-reversal symmetry (TRS). This Hall current is dissipationless, contrary to a longitudinal current, which breaks TRS through energy dissipation. \textcolor{black}{In the QHE, the TRS is broken by an external magnetic field \cite{TKNN} or internal magnetic fluxes, like in the Haldane model \cite{Haldane}. The QHE states, also called Chern insulators, are characterized by a non-zero Chern number, which indexes the topological phases of the system. }

\textcolor{black}{Chern insulators only exist in two dimensions and rely on time-reversal symmetry breaking. However, another topological class can be present even when TRS is preserved.} In the so-called quantum spin Hall effect (QSHE) \cite{KaneMele}, time-reversal symmetry is preserved due to the absence of external magnetic fields. Spin-orbit coupling generates an effective magnetic field that acts upward on the spin-up and downward on the spin-down electrons. As a result, electrons with opposite spins move in opposite directions (in separate conducting channels) at the edges of the sample. There is no net flow of charge, only a net spin current. In the QSHE, the spin current is even under the TRS, so $\sigma_{xy} \neq 0$ is consistent with this symmetry. In the top edge of a sample with a strip geometry, spin-up electrons move in one direction (say, right), spin-down electrons move in the opposite direction (left), and vice-versa at the bottom edge. \textcolor{black}{Systems supporting the QSHE have null a Chern number but can be indexed by a two-valued topological integer. They are often called $\mathbb{Z}_2$ topological insulators. The edge currents are protected by time-reversal symmetry.} Generally, if an electron backscatters, its spin is flipped, and time reversal symmetry is broken. But there will be no backscattering if TRS is present (as it happens when there is a non-magnetic impurity). Time-reversal symmetry is always present in $\mathbb{Z}_2$ topological insulators, so the edge modes are robust against backscattering. Topological systems have a topologically nontrivial band structure. Its bulk is insulating, and its surface/edge is a topologically protected conductor \cite{Phillips2012-jg,Teixeira_Pires2019-tx}.
    
\textcolor{black}{We can find topological states not only in electronic systems but also in ordered magnetic lattices, where quantized perturbations in the magnetic order, the so-called magnons, are the spin carriers.} Topological two-dimensional magnets display a finite magnon Hall conductivity that can be used in magnonic nanodevices. They can also show topologically protected edge modes, which are robust against structural or magnetic disorder and can form ideal waveguides for long-range magnon transport \cite{Ghader2021}. An essential property of noninteracting topological magnons is that they can propagate for a long time without dissipation. They do not interact strongly with other degrees of freedom and are, for this reason, adequate to be used in spintronics. The quantum Hall effect requires high magnetic fields, which limits its use in technology applications, whereas magnon transport does not require an external magnetic field. Note that in the QSHE, spins are carried out by the electrons. In magnets, the spin of the atoms is fixed, and what is carried out is the spin of the magnons (non-localized excitations): we have a magnonic spin current. Magnons can mediate various Hall-like transport phenomena in both ferromagnets (FM) and antiferromagnets (AF). For example, a transverse spin current appears in response to a magnetic field gradient (which plays the role of an electric field in electronic systems), or a transverse heat current, mediated by magnons, manifests in response to a thermal gradient. The spin Nernst effect is a spin current caused by a temperature gradient. The Nernst-Ettinghausen effect is a heat current in response to a magnetic field gradient \cite{Nakata2017}.

We should remark that electron and magnon systems are quite different. In the case of electrons, the energy levels are filled at zero temperature up to the Fermi energy (so we can have transport at zero temperature). We have conduction and valence bands, and conductors and insulators have different properties. For magnons, there is no Fermi energy. We have zero magnons at zero temperature and, therefore, no transport. We can have transitions from a filled band to another one since the Pauli exclusion principle does not apply here. Another difference from fermionic systems is that we can have a magnon spin Hall effect (spin current in response to a magnetic field gradient) even when the Chern number is zero. A topological ferromagnet (or antiferromagnet) is usually called a topological magnon insulator, although the term ``insulator" has no physical meaning for magnons. Here, we generically call magnon Hall effect the transverse current originated by a longitudinal field gradient. The effect can be topological when the Chern number differs from zero or geometrical when it vanishes. \textcolor{black}{As shown by Liu et al. \cite{Liu_arXiv}, Chern numbers in collinear antiferromagnets are expected to be zero. Some special in-plane magnetic configuration is required for a non-null Chern number.} Yet, it is possible to have topologically protected edge modes even when the Chern number is zero due to an effective time-reversal symmetry. That is the case of the pure spin Nernst effect in the brick-wall lattice studied in this paper.

    Transport phenomena in magnetic systems are essential tools for studying magnetic excitations and fluctuations. These techniques have become available as a probe due to the development of experimental methods in the context of spintronics, and in the last decades, there has been a great interest in studying topological effects in spin models [7-91]. Antiferromagnets exhibit ultrafast dynamics and are robust against reasonably large external fields. They are a competitive alternative to ferromagnets to store and manipulate information \cite{Bonbien2021}, as topological ferromagnets present drawbacks such as strong magnetic disturbances and low mobility due to high magnetization. \textcolor{black}{Topological transport in AF magnets is a developing field and has been investigated in several contexts: collinear paramagnets \cite{Zhang2018}, non-collinear and non-coplanar magnets \cite{Li2020,Kim2019,Bhowmick2020}, and dimmer systems \cite{Romhnyi2015,McClarty2017,Bhowmick2021}. Also, topological transport relies on a system's property called Berry curvature, which can be induced and controlled by several mechanisms, like electromagnetic waves \cite{Owerre_laser} or crystal chirality \cite{Mejkal2020}. Topological magnonics is a fast-evolving field. In this scenario, it is essential to study antiferromagnetic topological insulators in all kinds of lattices and compare their thermomagnetic properties.}
     
     This paper is organized as follows. In Section 2, we give a brief introduction to the magnon Hall effect following Ref. \cite{Nakata2017}. In Section 3, we study the transport of spins starting from the Kubo formula and show how to relate the Hall conductivity to the Berry curvature. Section 4 presents a generalized Boguliubov transformation to diagonalize the Hamiltonian of antiferromagnets. The Union Jack lattice is introduced in Section 5. In Section 6, we present a brief discussion of symmetries. In Section 7, the transport coefficients are introduced. The brick-wall lattice is investigated in Section 8. In Section 9, we summarize our results. Finally, in Appendix A, we treat the contribution of four operator terms using a self-consistent mean-field spin wave theory, and in Appendix B, we derive an expression for the Berry curvature from the eigenvalues of a general antiferromagnetic Hamiltonian.

\section{Magnon Hall effect}

As discussed in the following, one can make a reasonable (but not entirely equivalent) correspondence between the electronic spin degrees of freedom in a topological insulator with the two sublattice degrees of freedom in an antiferromagnet. In an analogy to TIs, we can study the topological properties of antiferromagnets and establish a bosonic version of the QSHE where the spin carriers are not electrons but magnons. In this section, we follow Ref. \cite{Nakata2017} and give a qualitative description of what happens in some topological antiferromagnets.

Based on the symmetry of Maxwell's equations, Aharonov and Casher \cite{AharonovCasher1984} considered the interaction between a particle's magnetic dipole moment and an electric field and introduced a dual to the Aharonov-Bohm effect, as we show in the following.

Suppose a two-dimensional ferromagnet in the $xy$-plane is subjected to a spatially varying electric field $\mathbf{E}(\mathbf{r})$. Through the Aharonov-Casher effect, $\mathbf{E}(\mathbf{r})$ couples to magnons' magnetic dipole moment $g\mu_{B}\mathbf{\hat{z}}$. A moving magnetic dipole interacts with the external electric field
and acquires an Aharonov-Casher phase due to hopping between the sites given by

\begin{equation}
\theta_{ij}=\frac{g\mu_{B}}{\hbar c^{2}}\int_{x_{i}}^{x_{j}}d\mathbf{r}%
\cdot\left(  \mathbf{E}(\mathbf{r})\times\mathbf{\hat{z}}\right)  .
\end{equation}

The exchange Hamiltonian in the presence of the AC effect can be
written as \cite{Nakata2017,Nakata2021}:

\begin{equation}
H=-\sum_{\left\langle i,j\right\rangle }\frac{J_{ij}}{2}\left[ \left(
S_{i}^{+}S_{j}^{-}e^{i\theta_{ij}}+S_{i}^{-}S_{j}^{+}e^{-i\theta_{ij}}\right)
+S_{i}^{z}S_{j}^{z}\right]  \label{AC_Hamiltonian}%
\end{equation}

where $\theta_{ij}$ is the AC phase the magnetic dipole moment,
associated with the spin along the z-direction, acquires when it hops between
neighboring sites\ $x_{i}$ and $x_{j}$.

The Hamiltonian of a ferromagnet with Dzyaloshinskii-Moriya (DM) interaction with the DM vector
given by $\mathbf{D}_{ij}=\pm D \mathbf{\hat{z}}$ can be written as an expression analogous to
Eq. (\ref{AC_Hamiltonian}), even without the presence of an electric field $\mathbf{E(r)}$ \cite{LeeHanLee2015}, as we show in the following.

A Hamiltonian with FM exchange and Dzyaloshinskii-Moriya interaction between near-neighbor sites is written as:

\begin{align}
H&=-J\sum_{\left\langle i,j\right\rangle }\mathbf{S}_{i}\cdot\mathbf{S}%
_{j}+\sum_{\left\langle i,j\right\rangle }\mathbf{D\cdot S}_{i}\times
\mathbf{S}_{j} \nonumber \\
&=-J\sum_{\left\langle i,j\right\rangle }\mathbf{S}_{i}%
\cdot\mathbf{S}_{j}\pm D\sum_{\left\langle i,j\right\rangle }\mathbf{\hat
{z}\cdot S}_{i}\times\mathbf{S}_{j}%
\end{align}

Using the ladder operators, we can develop this as:

\begin{align}
H  &  =-J\sum_{\left\langle i,j\right\rangle }\left(  S_{i}^{x}S_{j}^{x}%
+S_{i}^{y}S_{j}^{y}+S_{i}^{z}S_{j}^{z}\right)  \pm D\sum_{\left\langle
i,j\right\rangle }\left(  S_{i}^{x}S_{j}^{y}-S_{i}^{y}S_{j}^{x}\right) \nonumber  \\
&  =-\frac{J}{2}\sum_{\left\langle i,j\right\rangle }\left(  S_{i}^{+}%
S_{j}^{-}+S_{i}^{-}S_{j}^{+}\right)  \pm\frac{iD}{2}\sum_{\left\langle
i,j\right\rangle }\left(  S_{i}^{+}S_{j}^{-}-S_{i}^{-}S_{j}^{+}\right)
+J\sum_{\left\langle i,j\right\rangle }S_{i}^{z}S_{j}^{z} \nonumber \\
&  =-\sum_{\left\langle i,j\right\rangle }\left(  \frac{J\mp iD}{2}S_{i}%
^{+}S_{j}^{-}+\frac{J\pm iD}{2}S_{i}^{-}S_{j}^{+}\right)  +J\sum_{\left\langle
i,j\right\rangle }S_{i}^{z}S_{j}^{z}%
\end{align}

We can define a phase $\theta=\tan^{-1}\left(  D/J\right)  $, so we have:

\begin{equation}
\sin\theta=\frac{D}{\sqrt{J^{2}+D^{2}}}, \, \, \, \, \cos\theta=\frac
{J}{\sqrt{J^{2}+D^{2}}}
\end{equation}

Also note that $J\pm iD=\sqrt{J^{2}+D^{2}} \, e^{\pm i\theta}$. Hence, The
Hamiltonian can be written as

\begin{equation}
H=-\frac{\tilde{J}}{2}\sum_{\left\langle i,j\right\rangle }\left(  e^{\mp
i\theta} \, S_{i}^{+}S_{j}^{-}+e^{\pm i\theta} \, S_{i}^{-}S_{j}%
^{+}\right)  +J\sum_{\left\langle i,j\right\rangle }S_{i}^{z}S_{j}^{z}
\end{equation}

with $\tilde{J}=\sqrt{J^{2}+D^{2}}$. That has the same form as Eq. (\ref{AC_Hamiltonian}), showing that the
DM vector $\mathbf{D}$ plays the role of a vector potential for magnons, and an \textcolor{black}{external
electric field} is not necessary to induce the effects to be described below.
However, for simplicity, we will consider that the source of the AC effect is
a \textcolor{black}{(spatially varying) static electric field}, as it is usually done when presenting
this subject. For an isotropic antiferromagnet (for instance, in a square
lattice) with spin-rotational symmetry around the z-axis and in the linear
spin wave approximation, up and down magnons are completely decoupled, and the dynamics of magnons is described as the combination of two independent copies of the dynamics of magnons in a ferromagnet for each mode [4,5]. To proceed with the discussion, we will consider a Hamiltonian with degenerate up and down magnons. 

In the absence of the AC term, a magnetic field
gradient $\partial_{x}B$ along the x-axis gives origin to a force $F_{\sigma
}=\sigma g\mu_{B}\partial_{x}B$, where $\sigma=-1$ for the ferromagnet and
$\sigma=\pm1$ for the antiferromagnet. That drives antiferromagnetic magnons
in the bulk with opposite magnetic moments to flow in opposite $\pm \mathbf{\hat{x}}$ directions (since the directions of the forces are opposite for the two magnons modes),
generating a longitudinal spin current without a net heat current. On the
other hand, when subjected only to a thermal gradient $\partial_{x}T$, both spin-up
and spin-down magnons flow in the same direction. The spin current vanishes,
but there is a net longitudinal heat current. That is the conventional
transport behavior of an antiferromagnetic magnonic system (see first column of Table \ref{table}).

To study transversal (Hall) transport in antiferromagnets, we analyze
magnons in an electric field $\mathbf{E}\left(  \mathbf{r}\right)  $.
We must replace the momentum operator $\mathbf{p}$ by $\mathbf{p}+\sigma g\mu
_{B}\mathbf{A}/c$, where \cite{AharonovCasher1984}:

\begin{equation}
\mathbf{A}\left(  \mathbf{r}\right)  =\frac{1}{c}\mathbf{E}\left(  \mathbf{r}\right)
\times\mathbf{\hat{z}} \label{def_A}%
\end{equation}

Hence, the Hamiltonian in the low-energy dynamics and in the continuum
approximation (that is, the Hamiltonian of a gas of noninteracting magnons) is
given by $H=\sum_{\sigma=\pm}H_{\sigma}$, with:%

\begin{equation}
H_{\sigma}=\frac{1}{2m}\left(  \mathbf{p}+\frac{\sigma g\mu_{B}}{c}\mathbf
{A}\right)  ^{2}-\sigma g\mu_{B}B \label{hamiltonian_EM}%
\end{equation}

where $m$ is the effective magnon mass. We can see that the Hamiltonian above
is formally identical to the one from a charged particle on a magnetic field,
with $\mathbf{A}$ playing the role of the vector potential and $\sigma g\mu_{B}$
being the coupling constant (instead of $e$).

If the \textcolor{black}{spatially varying static electric field has the form $\mathbf{E}\left(  \mathbf{r}\right)  =E\left(
-x/2,-y/2,0\right)$, where $E$ is a constant, from Eq. (\ref{def_A}) we obtain a
gauge potential $\mathbf{A}\left(  \mathbf{r}\right)  =\left(  E/c\right)  \left(
-y/2,+x/2,0\right)$, which gives} :

\begin{equation}
\mathbf{\nabla}\times\mathbf{A}=\frac{E}{c}\mathbf{\hat{z}}, \label{rotA}%
\end{equation}

and we can see that $E$ plays the role that, in charged particles, is played
by magnetic fields.

From the canonical equation of motion, we find that the force acting on
magnons is

\begin{equation}
\mathbf{F}_{AC}=\sigma g\mu_{B}\left[  \mathbf{\nabla}B-\frac{\mathbf{v}}{c}%
\times\left(  \mathbf{\nabla}\times\mathbf{A}\right)  \right]  . \label{force}%
\end{equation}

Using (\ref{rotA}), we can write%

\begin{equation}
\mathbf{F}_{AC}=\sigma g\mu_{B}\left[  \mathbf{\nabla}B-\frac{\mathbf{v}}{c^{2}}\times
E\mathbf{\hat{z}}\right]  , \label{force_2}%
\end{equation}

and again, comparing $\mathbf{F}_{AC}$ to the Lorentz force in charged particles
$\left(  \mathbf{F}_{e}=e\left[  \mathbf{E}+\mathbf{v}\times\mathbf{B}\right]  \right)  $,
we note the parallel between electronic and magnonic transport in the presence
of EM fields: the roles that electric and magnetic fields have in electronic
systems is played, respectively, by \textcolor{black}{$\mathbf{\nabla
}B$ and $-E\mathbf{\hat{z}}/c^2$ in magnonic systems}.

The velocity $\mathbf{v}$ is constituted of two parts: the cyclotron velocity
$\mathbf{v}_{c}$ and the drift velocity $\mathbf{v}_{d}$. The drift velocity is the velocity of the guiding center, meaning that in the drifting frame, the velocity comes only from the cyclotron motion. Supposing that $\frac
{d}{dt}\mathbf{v}_{d}=0$, we find $\mathbf{v}_{d}\times\mathbf{\hat{z}}=\left(
c^{2}/E\right)  \mathbf{\nabla}B$ \cite{Nakata2017}. If the magnetic field gradient is applied
along the x-axis ($\mathbf{\nabla}B=\partial_{x}B \, \mathbf{\hat{x}}$), we get $\mathbf{v}_{d}=\left(  0,\left(  c^{2}/E\right)
\partial_{x}B,0\right)$, which is independent of $\sigma$ and perpendicular to $\partial_{x}B \, \mathbf{\hat{x}}$. Thus, all magnons (with spin up and down) flow in the same $\mathbf{\hat{y}}$ direction when subjected to
electric and magnetic field gradients. The transversal spin current vanishes
in the bulk, while the heat current in nonzero. On the other hand, a thermal gradient generates helical magnon Hall transport in the bulk where up and down magnons flow in opposite directions. Since magnons of up and down spins convey down and up spins, respectively, a nonzero spin current appears while the total heat current cancels out (see the second column of Table \ref{table}). A more detailed discussion about the thermomagnetic properties can be done using Onsager coefficients, which relate thermal and magnetic field gradients with current and heat densities in the bulk \textcolor{black}{\cite{Mahan2012-df}}.

That qualitatively describes the Hall transport of magnons in the bulk of an
AF magnet, which is generated by coupling magnons with a \textcolor{black}{static and spatially varying electric field} through the Aharonov-Casher effect. The
thermomagnetic properties of AF magnets are summarized in Table \ref{table}. We stress
that these conclusions were made in the assumption that we are in the linear
response regime, where sufficiently low temperature and applied magnetic
fields make the energy bands almost degenerate. If the band degeneracy is lifted, as in the
case of the Union Jack lattice studied in Section 5, both spin and heat Hall currents can be simultaneously different from zero.

\textcolor{black}{Even though the formalism above requires static electric fields or DM vectors, it is known that the same result can be achieved with an oscillating electromagnetic field \cite{Owerre_laser}. That has the advantage that electromagnetic waves can be easily tuned. The Floquet theory enables one to transform a time-dependent model into an effective static model governed by the so-called Floquet Hamiltonian. As a result, we get a synthetic tunable intrinsic DMI in quantum magnets without an inversion center. This can be achieved with the application of a circularly polarized laser with a dominant oscillating electric field component perpendicular to the magnetic 2D material. Another advantage of oscillating fields is that they amplify topological magnons via dipolar coupling \cite{dip1,dip2}. This interaction contributes to the complex off-diagonal term $f_k$ of a magnon Hamiltonian (See Eq. (\ref{hamilt_matrix})), which is related to the Berry curvature. That term is associated with the DMI in the Union Jack lattice, while in the brick-wall lattice it is associated with a complex structure factor. The interaction with an oscillating electric field can amplify these terms and enhance transverse transport.}

Regarding the edge properties of the system, we see from \textcolor{black}{the discussion under Eq. (\ref{force_2}) that
the static drift velocity $\mathbf{v}_{d}=\left(  0,\left(  c^{2}/E\right)
\partial_{x}B,0\right)$} vanishes in the bulk if \textcolor{black}{the magnetic field gradient is zero}. However, each magnon
still performs a cyclotron motion in opposite directions due to the \textcolor{black}{static electric
field $\mathbf{E}(\mathbf{r})$}. So in the presence of a thermal gradient, we have helical edge magnon states: up and down
magnons propagate along the edges of the sample in opposite directions. That
is a bosonic analogue of the QSHE for electrons. Hence, an AF magnet in
the presence of the Aharonov-Casher effect can be considered a magnonic
topological insulator.

All these considerations were made assuming the system has a N\'{e}el ground state. In this work, we added an easy-axis anisotropy that ensures the magnetic Néel order is along the $\mathbf{\hat{z}}$ direction.

\begin{table}
\[%
\begin{tabular}
[c]{|c|c|c|}
\hline
& \begin{tabular}
[c]{c}%
Trivial transport\\
(longitudinal)
\end{tabular} &
\begin{tabular}
[c]{c}%
Hall transport\\
(transverse)
\end{tabular}
\\
\hline
Magnetic field gradient &
\begin{tabular}
[c]{l}%
Spin current $\neq$ $0$\\
Heat current $=0$%
\end{tabular}
&
\begin{tabular}
[c]{l}%
Spin current $=$ $0$\\
Heat current $\neq$ $0$%
\end{tabular}
\\
\hline
Thermal gradient &
\begin{tabular}
[c]{l}%
Spin current $=$ $0$\\
Heat current $\neq$ $0$%
\end{tabular}
&
\begin{tabular}
[c]{l}%
Spin current $\neq$ $0$\\
Heat current $=0$%
\end{tabular}\\
\hline
\end{tabular}
\]

\caption{\label{table} Magnon transport induced by a magnetic field and thermal gradients in the topologically trivial and nontrivial bulk of the antiferromagnet described by Hamiltonian (\ref{hamiltonian_EM}).}
\end{table}

\section{Transport}

An electric field produces an electric current, and a magnetic field gradient
gives rise to a spin current. Within the linear response theory, the Kubo
formula for conductivity is given by \cite{Mahan2012-df,Kubo1985-jv,spinmodels_pires}:

\begin{equation}
\sigma_{\alpha\beta}\left(  \mathbf{q},\omega\right)  =-\frac{i}{V}\sum_{n,m}%
\sum_{k}\left(  \frac{n_{n,k}-n_{m,k+q}}{E_{n,k}
-E_{m,k+q}}\right)  \frac{\left\langle \mathbf
{k},n\right\vert J_{\alpha,k} \left\vert \mathbf{k}+\mathbf
{q},m\right\rangle \left\langle \mathbf{k}+\mathbf{q},m\right\vert J_{\beta,k+q} \left\vert \mathbf{k},n\right\rangle }{\omega
+i\eta+E_{n,k}-E_{m,k+q}}%
\label{kubo}
\end{equation}

where $V$ is the volume; $J_{\alpha,k}$ is the current in
the $\alpha$ direction at point $\mathbf{k}$ of the Brillouin zone; $E_{n,k}$ is the energy of th $n$-th band and $n_{n,k}$ is the Fermi-Dirac distribution function for fermions or the Bose distribution
function for bosons.
The states\ $\left\vert \mathbf{k},n\right\rangle $ in Eq. (\ref{kubo}) are the exact
eigenstates of the Hamiltonian and $E_{n,k}$ are the
exact energy levels. However, in the spin wave approximation we use
one-particle states of the noninteracting Hamiltonian.

The $\mathbf{q}=0$ longitudinal magnon conductivity is given by:

\begin{equation}
\sigma_{xx} \left(\omega\right)  =-\frac{i}{V}\sum_{n,m}%
\sum_{k}\left(  \frac{n_{n,k}-n_{m,k}}{E_{n,k}
-E_{m,k}}\right)  \frac{\left\langle \mathbf
{k},n\right\vert J_{x,k} \left\vert \mathbf{k},m\right\rangle \left\langle \mathbf{k},m\right\vert J_{x,k} \left\vert \mathbf{k},n\right\rangle }{\omega
+i\eta+E_{n,k}-E_{m,k}}% 
\label{xx}
\end{equation}

The index $n$ is identified with a band. We can perform the thermodynamic limit from discrete to continuous $\mathbf{k}$ (where $\mathbf{k}$ runs in each band). Besides the interband (regular) term, we have the additional intraband term (inside each band) where $m \rightarrow n$. Using

\begin{equation}
\frac{n_{n,k}-n_{m,k}}{E_{m,k}-E_{n,k}}=\frac{1}{\Delta}\left[ n_{n,k}- \left( n_{n,k} + 
\frac{\partial n_{n,k} }{\partial E_{n,k}} \right) \Delta \right]=\frac{\partial n_{n,k} }{\partial E_{n,k}} 
\end{equation}

where $\Delta =E_{m}-E_{n} $, we get

\begin{equation}
\sigma_{xx}^{intraband}\left(\omega\right)= \frac{1}{\omega + i\eta} \sum_{k}
\frac{\partial n_{n,k} }{\partial E_{n,k}} \left\langle \mathbf
{k},n\right\vert J_{x,k} \left\vert \mathbf{k},n\right\rangle \left\langle \mathbf{k},n\right\vert J_{x,k} \left\vert \mathbf{k},n\right\rangle 
\end{equation}

The real part of the dynamical spin conductivity $\sigma_{xx}\left(
\omega\right)$ is written as

\begin{equation}
Re \  \sigma_{xx}\left(  \omega\right)  =D\left(  T\right)
\delta\left(  \omega\right)  +\sigma^{reg}\left(  \omega\right)
\end{equation}

Where the coefficient of the zero frequency $\delta$ function contribution is called the Drude weight, and it is given by

\begin{equation}
D\left(  T\right)  = \pi \sum_{k}\frac{\partial n_{n,k} }{\partial E_{n,k}} \left\langle \mathbf
{k},n\right\vert J_{x,k} \left\vert \mathbf{k}%
,n\right\rangle \left\langle \mathbf{k},n\right\vert J_{x,k}
\left\vert \mathbf{k},n\right\rangle
\end{equation}

This term is interpreted as the contribution of thermally excited particles propagating ballistically without interacting with other particles.

The regular part of the longitudinal conductivity $\sigma_{xx}^{reg}$ is given by:

\begin{equation}
\sigma_{xx}^{reg} \left(\omega\right)  =-\frac{1}{V}\sum_{n,m}%
\sum_{k}\left(  \frac{n_{n,k}-n_{m,k}}{E_{n,k}
-E_{m,k}}\right) \left\langle \mathbf{k},n\right\vert J_{x,k} \left\vert \mathbf{k},m\right\rangle \left\langle \mathbf{k},m\right\vert J_{x,k} \left\vert \mathbf{k},n\right\rangle \delta \left[ \omega- \left( E_{n,k}
-E_{m,k} \right) \right]% 
\end{equation}

The transverse static spin conductivity in a two-dimensional system, called
the spin Hall conductivity, is

\begin{equation}
\sigma_{xy}=\sigma_{xy}\left(  0,0\right)  =\frac{i}{A}\sum_{k}\sum_{m\neq
n} \frac{ \left(  n_{n,k}-n_{m,k} \right)  \left\langle \mathbf{k},n\right\vert
J_{x,k} \left\vert \mathbf{k},m\right\rangle \left\langle
\mathbf{k},m\right\vert J_{y,k} \left\vert \mathbf
{k},n\right\rangle }{\left[  E_{n,k} -E_{m,k} \right]  \left[  E_{n,k} -E_{m,k}  +i\eta\right]  },
\label{sigma_1}%
\end{equation}

where A is the area for a two-dimensional system. Using the formula for the spin current \cite{Han2017}:

\begin{equation}
J_{\alpha}=g\mu_{B}\frac{\partial H_{k}}{\partial k_{\alpha}},
\end{equation}

and defining%

\begin{equation}
\Omega_{\alpha\beta}^{n}\left(  \mathbf{k}\right)  \equiv i\sum_{m\neq n}%
\frac{\left\langle \mathbf{k},m\right\vert \partial H_{k}/\partial k_{\alpha
}\left\vert \mathbf{k},n\right\rangle \times\left\langle \mathbf{k},n\right\vert
\partial H_{k}/\partial k_{\beta}\left\vert \mathbf{k},m\right\rangle }{\left(
E_{m,k}  -E_{n,k} \right)  ^{2}}
\label{berry_c_1}%
\end{equation}

the spin Hall conductivity $\sigma_{xy}$ can be written as

\begin{equation}
\sigma_{xy}=\left(  g\mu_{B}\right)  ^{2}\sum_{\lambda}\int\limits_{BZ}%
\frac{dk_{x}dk_{y}}{\left(  2\pi\right)  ^{2}}n_{k}^{\lambda}\Omega
_{xy}^{\lambda}(  \mathbf{k})  \label{sigma_2}%
\end{equation}

The term $\Omega_{xy}^{\lambda}(  \mathbf{k})  $ is the Berry
curvature of the $\lambda$-th band (and from now on, the superscript indexes the bands). It is this term that gives rise to a
transverse motion of magnons and leads to a Hall response. If $H_{k}$ is real,
there is always a choice to have real eigenvectors (unless there is a band
degeneracy), which renders $\Omega_{xy}^{\lambda}(  \mathbf{k})  =0$.
Therefore, the existence of an imaginary part of the Hamiltonian is a
necessary condition to have a non-null Berry curvature. In the Union Jack lattice studied in Section 5, the imaginary part comes from the Dzyaloshinskii-Moriya term. In the brick-wall lattice studied in Section 8, it comes from a complex structure factor. Also, the gap should not vanish.

We can insert $\sum\limits_{m^{\prime}}\left\vert \mathbf{k},m^{\prime
}\right\rangle \left\langle \mathbf{k},m^{\prime}\right\vert = I$  into Eq.
(\ref{berry_c_1}) and obtain

\begin{equation}
\left\langle \mathbf{k},n\right\vert \frac{\partial}{\partial k_{\alpha}%
}\left\vert \mathbf{k},m^{\prime}\right\rangle \left\langle \mathbf{k},m^{\prime
}\right\vert H_{k}\left\vert \mathbf{k},m\right\rangle
\end{equation}

Renaming the eigenvectors as $u_{nk}$, it is easy to show that we can write

\begin{equation}
\Omega_{\alpha\beta}^{n}\left(  \mathbf{k}\right)  =i\left(  \frac{\partial
u_{nk}^{\dagger}}{\partial k_{\alpha}}\frac{\partial u_{nk}}{\partial
k_{\beta}}-\frac{\partial u_{nk}^{\dagger}}{\partial k_{\beta}}\frac{\partial
u_{nk}}{\partial k_{\alpha}}\right)  \text{ \ \ , \ \ \ }\left(  \alpha
,\beta=x,y\right)  \label{berry_c_final}%
\end{equation}

or

\begin{equation}
\Omega_{xy}^{n}\left(  \mathbf{k}\right)  =i\sum_{\alpha\beta}\varepsilon
_{\alpha\beta}\left(  \frac{\partial u_{nk}^{\dagger}}{\partial k_{\alpha}%
}\frac{\partial u_{nk}}{\partial k_{\beta}}\right)  \text{ \ \ \ ,
\ \ \ }\left(  \alpha,\beta=x,y\right)
\end{equation}

where $n=1,2,\ldots N$ ($N$ is the number of bands) and $\varepsilon_{\alpha \beta}$ is the antisymmetric tensor. The integral of the Berry
curvature in the Brillouin zone, associated with band $n$, is the (first)
Chern number $C^{(n)}$, which is an integer and temperature-independent so it
cannot change its value continuously. Note that the Berry curvature measures
the phase accumulated by the ground state eigenfunctions when evolving
in the Brillouin zone. It is a local geometric object since it
explicitly depends on $\mathbf{k}$. On the other hand, the Chern number is a
global topological index. The sum of the Chern numbers over all bands is zero. The gap between neighboring bands has to be different from zero for systems with a non-null Chern
number. However, a non-null gap does not imply a $C\neq 0.$

The Chern number is
a topological invariant. That means that if we deform the energy bands (by varying the Hamiltonian parameters) without closing the gap, the Chern number will not change. Suppose that $C=1$ in the
bulk of a material. We have, naturally, $C=0$ in the vacuum. In the very edge of the material, the Chern number has to change from $C=1$ to $C=0$, i.e., a topological phase transition. The localized states at the edge have to show gapless energy bands to allow that transition. Therefore, we have gapped bulk states and gapless edge states. We do not need to perform theoretical calculations in a finite sample to find out if we have gapless edge modes: the calculation of the Chern number in
the bulk is sufficient.

In the case of fermions, if there exists an energy gap between the upper and
lower bands, and the lower band is fully filled, that is $E_{-,k}%
<E_{F}<E_{+,k}$ (where $E_{F}$ is the Fermi energy), then $f_{+,k}=0$ and
$f_{-,k}=1$ at zero temperatures. Thus, the Hall conductivity of filled bands is
given by

\begin{equation}
\sigma_{xy}=e^{2}\sum_{n=filled}C^{\left(  n\right)  }%
\end{equation}

and is quantized. In the case of a magnon current, there are no filled bands, the Bose
factor vanishes at zero temperature, and the current is not quantized.
However, as mentioned above, a nonzero Chern number implies topologically
protected edge states. The correspondence between the Chern number and the
number of gapless edge states still holds for magnons.

\section{Generalized Bogoliubov transformation}

Here we present a formalism to diagonalize antiferromagnetic Hamiltonians. We
start with a general quadratic bosonic Hamiltonian written in a matrix form as

\begin{equation}
H=\sum_{k}\psi_{k}^{\dagger}H_{k}\psi_{k} \label{hamilt_1}%
\end{equation}

where%

\begin{equation}
\psi_{k}^{\dagger}=\left(  b_{1k}^{\dagger}\,,...,\,b_{Nk}^{\dagger
}\,,\,b_{1,-k}\,,\,...\,,b_{N,-k}\right)  \label{basis_old}%
\end{equation}

We must diagonalize the matrix $H_{k}$ to obtain the magnon spectrum. We need
to find a transformation matrix $T_{k}$ from a new basis $\varphi_{k}$ to the
old basis: $\psi_{k}=T_{k}\varphi_{k}.$

Let $\alpha_{mk}\,(  \alpha_{mk}^{\dagger})  $ be the eigenstates of
$H_{k}$, and write

\begin{equation}
\varphi_{k}^{\dagger}=\left(  \alpha_{1k}^{\dagger},...,\alpha_{Nk}^{\dagger
},\alpha_{1,-k},...,\alpha_{N,-k}\right)  \label{basis_new}%
\end{equation}

After transformation, the components of $\varphi_{k}$ must satisfy the same
commutation relation as $\psi_{k}$, that is

\begin{equation}
\left[  \varphi_{ik},\varphi_{ik}^{\dagger}\right]  =\left[  \psi_{ik}%
,\psi_{ik}^{\dagger}\right]  =\eta_{ij}%
\end{equation}

where%

\begin{equation}
\eta=\left(
\begin{array}
[c]{cc}%
I_{N\times N} & 0\\
0 & -I_{N\times N}%
\end{array}
\right)
\end{equation}

Here, $I_{N\times N}$ is a unit matrix. We must ensure that the commutation
relations of the bosonic operators are conserved by the transformation $T_{k}%
$. The matrix $T_{k}$ is chosen such that the Hamiltonian (\ref{hamilt_1}) can
be written as

\begin{equation}
H=\sum_{k}\phi_{k}^{\dagger}T^{\dagger}H_{k}T\phi_{k}=\sum_{k}\phi
_{k}^{\dagger}\tilde{H}_{k}\phi_{k} \label{hamilt_2}%
\end{equation}

where $\tilde{H}_{k}$ is diagonal:

\begin{equation}
\tilde{H}_{k}=T^{\dagger}H_{k}T=\left(
\begin{array}
[c]{cccc}%
\omega_{1k} & 0 & ... & 0\\
0 & \omega_{2k} & ... & 0\\
... & ... & ... & ...\\
0 & 0 & ... & \omega_{N,-k}%
\end{array}
\right)
\end{equation}

In terms of the basis (\ref{basis_new}), Eq. (\ref{hamilt_2}) reads

\begin{equation}
H=\sum_{k}\sum_{n=1}^{N}\omega_{nk}\left(  \alpha_{nk}^{\dagger}\alpha
_{nk}+\frac{1}{2}\right)
\end{equation}

To ensure that the new operators $\alpha_{nk}$ satisfy the bosonic algebra,
the matrix $T_{k}$ must fulfill the condition:

\begin{equation}
T_{k}\eta T_{k}^{\dagger}=\eta\label{paraunitary}%
\end{equation}

A matrix $T_{k}$ satisfying condition (\ref{paraunitary}) is referred to as paraunitary.

The eigenvalues of $H_{k}$ are obtained by the diagonalization of the matrix

\begin{equation}
K_{k}\equiv\eta H_{k}%
\end{equation}

The matrix $K_{k}$ is non-Hermitian, but it can still be diagonalized by
different left and right eingenstates with corresponding real eigenvalues. We
should remember that a right eigenvector of a matrix $A$ is a column matrix
$u$ that satisfies $Au=\lambda u.$ A left eigenvector of $A$ is a row matrix
$v$ that satisfies $vA=\lambda v.$

The matrix $T_{k}$ consists of all the eigenvectors of $K_{k}$:

\begin{equation}
T=\left[  V\left(  \omega_{1}\right)  ,...,V\left(  \omega_{N}\right)
,V\left(  -\omega_{1}\right)  ,...,V\left(  -\omega_{N}\right)  \right]
\end{equation}

with the eigenvectors $V$ ordered as

\begin{equation}
V^{\dagger}\left(  \omega_{i}\right)  \eta V\left(  \omega_{i}\right)
=1\text{ \ \ , \ \ \ }V^{\dagger}\left(  -\omega_{i}\right)  \eta V\left(
-\omega_{i}\right)  =-1 \label{norm}%
\end{equation}

for each set $\left(  V\left(  \omega_{i}\right)  ,V\left(  -\omega
_{i}\right)  \right)  $. The matrix $T_{k}$ diagonalizes $K_{k}$ and $H_{k}$ simultaneously%

\begin{align}
T_{k}^{-1}K_{k}T_{k}  &  \equiv\tilde{K}_{k}=diag\left[  \omega_{1}%
,...,\omega_{N},-\omega_{1},...,-\omega_{N}\right] \\
T_{k}^{\dagger}H_{k}T_{k}  &  \equiv\tilde{H}_{k}=diag\left[  \omega
_{1},...,\omega_{N},\omega_{1},...,\omega_{N}\right]
\end{align}

From\ Eq. (\ref{norm}), we see there are two different normalizations for the
eigenvectors of $K_{k}.$ Also, all the eigenvalues are real and appear in
pairs $\pm\omega_{i}$. It is usual to refer to the bands with indices
$n=1,\ldots,N$ $\ (n=N+1,\ldots,2N)$ as particle (hole) bands \cite{Samajdar2019}. Particle
states have normalization $+1$, while hole states have normalization $-1$. For
more details, see Refs. \cite{Samajdar2019,Colpa1978}.

We also note that the Hamiltonian matrix $H_{k}$ satisfies the particle-hole
symmetry \cite{Matsumoto2014}:

\begin{equation}
H_{k}=\rho H_{-k}^{T}\rho;\text{ \ \ \ \ \ \ \ \ }\rho=\left(
\begin{array}
[c]{cc}%
0 & I_{N\times N}\\
I_{N\times N} & 0
\end{array}
\right)
\end{equation}

The procedure shown above establishes a specific order for the $b_{i}^{\dagger
}$ operators in $\psi_{k}^{\dagger}\ $\ (see Eq. (\ref{basis_old})). But sometimes
it is convenient to establish a different order. In general, we should write
$\eta$ as

\begin{equation}
\eta=\left(
\begin{array}
[c]{cccc}%
\sigma_{1} & 0 & 0 & 0\\
0 & \sigma_{2} & 0 & 0\\
0 & 0 & ... & 0\\
0 & 0 & 0 & \sigma_{2N}%
\end{array}
\right)
\end{equation}

where $\sigma_{i}=+1$ $\left(  -1\right)  $ if the correspondent operator in
$\psi_{k}^{\dagger}$ is $b_{k}^{\dagger}$ $\left(  b_{-k}\right)  $. That
shuffles the particle/hole eigenvectors in $T_{k}$, but it can be useful if
the corresponding matrix $H_{k}$ is block-diagonal in this basis, which is the
case of the Union Jack lattice.

In terms of the right and left eigenstates, we have:

\begin{equation}
\eta H_{k}\left\vert n,\mathbf{k}\right\rangle _{R}=\left(  E_{k}\right)
_{nn}\left\vert n,\mathbf{k}\right\rangle _{R},\text{ \ \ \ }\left\langle
n,\mathbf{k}\right\vert _{L}\eta H_{k}=\left(  E_{k}\right)  _{nn}\left\langle
n,\mathbf{k}\right\vert _{L}%
\end{equation}

where

\begin{equation}
\left\vert n,\mathbf{k}\right\rangle _{R}=T_{n,k},\text{ \ \ \ }\left\langle
n,\mathbf{k}\right\vert _{L}=T_{n,k}^{\dagger}\text{\ }%
\end{equation}

We will use only the right eigenstate and write $\left\vert n,\mathbf
{k}\right\rangle _{R}=\left\vert n,\mathbf{k}\right\rangle .$ The normalization
relation becomes%
\begin{equation}
\text{\ }\left\langle n,\mathbf{k}\right\vert \eta\left\vert m,\mathbf
{k}\right\rangle =\eta_{nm}%
\end{equation}

The Berry connection for a bosonic system is given by \cite{Shindou2013}

\begin{equation}
A_{\alpha}^{n}\equiv i\left[  \eta T_{k}^{\dagger}\eta\frac{\partial T_{k}%
}{\partial k_{\alpha}}\right]  _{nn} \label{berry_con}%
\end{equation}

Note that if the eigenstate of the $n$-th band is multiplied by a phase $e^{ i\theta (\mathbf{k})}$ that is a smooth function of $\mathbf{k}$, then (\ref{berry_con}) transforms as $A_{\alpha}^{n}\rightarrow A_{\alpha}^{n}-\partial_{k_{\alpha}}%
\theta(  \mathbf{k})$ under this phase change.

The Berry curvature is defined as

\begin{equation}
\Omega_{\alpha\beta}^{n}(  \mathbf{k})  \equiv\frac{\partial A_{\beta
}^{n}}{\partial k_{\alpha}}-\frac{\partial A_{\alpha}^{n}}{\partial k_{\beta}}%
\end{equation}

Using (\ref{berry_con}) we get \cite{Samajdar2019}

\begin{equation}
\Omega_{xy}^{n}(  \mathbf{k})  =i\sum_{\alpha\beta}\varepsilon
_{\alpha\beta}\left[  \eta\frac{\partial T_{k}^{\dagger}}{\partial k_{\alpha}%
}\eta\frac{\partial T_{k}}{\partial k_{\beta}}\right]  _{nn},\text{
\ \ }\alpha,\beta=x,y \label{berry_c}%
\end{equation}

The Berry curvature is gauge-invariant. Using Stoke's theorem, we
see that the integral of the Berry curvature over the Brillouin zone is zero
if $A_{\alpha}^{n}\left(  \mathbf{k}\right)  $ is a
smooth function across the zone. Only when it is impossible to parameterize
the eigenstates over the entire Brillouin zone with a single gauge choice we
get a nonzero Chern number. The sum of the Chern numbers over all particle and hole bands is individually zero \cite{Samajdar2019}.

In the Heisenberg picture, the operator $\psi_{k}\left(  t\right)
$ satisfies \cite{Wang2021}

\begin{equation}
\frac{d\psi_{k}}{dt}=\frac{i}{\hbar}\left[  \psi_{k}^{\dagger}H_{k}\psi
_{k},\psi_{k}\right]  =\frac{i}{\hbar}\left[  \psi_{k}^{\dagger},\psi
_{k}\right]  H_{k}\psi_{k}=-\frac{i}{\hbar}\eta H_{k}\psi_{k}%
\end{equation}

That is, $\psi_{k}(t)$ is the solution of a non-Hermitian Schr{\"o}dinger-like equation

\begin{equation}
i\hbar\frac{d\psi_{k}}{dt}=\eta H_{k}\psi_{k}%
\end{equation}

The non-Hermiticity modifies the inner product for the boson wave functions as
$\left\langle \phi_{a}|\phi_{b}\right\rangle =\phi_{a}^{\dagger}\eta\phi_{b}.$

\section{The Union Jack Lattice}

Thermal Hall conductivity in antiferromagnets is studied mainly in the
kagome \cite{Lu2019,Laurell2018,Mook2019} and honeycomb \cite{Cheng2016,Owerre8,Zyuzin2016} geometry, but it has already been investigated in the square \cite{Samajdar2019,Zhang2013}, checkerboard \cite{Pires2020,Pires2021} and variations of the Lieb lattices \cite{Le2019,deOliveira2023,Bhattacharya2019}. Using the Schwinger boson mean-field theory, Samajdar et al. \cite{Samajdar2019} extended the calculations to spin-liquids. With the intent to find novel antiferromagnetic systems where Hall-like transport is present, here we study a two-dimensional AF Union Jack lattice with the Hamiltonian given by:

\begin{align}
H  &  =J_{1}\sum_{\left\langle i,j\right\rangle }\mathbf{S}_{i}\cdot
\mathbf{S}_{j}+\sum_{\left\langle \left\langle i,i'\right\rangle \right\rangle
}J_{2,ii'}\left(  \mathbf{S}_{i}\cdot\mathbf{S}_{i'}+\left(  \lambda-1\right)
S_{i}^{z}S_{i'}^{z}\right)  +\nonumber\\
&  + D \sum_{\left\langle i,j\right\rangle } \nu_{ij} \mathbf{\hat{z}} \cdot  \mathbf{S}_{i} \times \mathbf{S}_{j} -A\sum_{i}\left(  S_{i}^{z}\right)  ^{2}
\label{hamilt}%
\end{align}

\begin{figure}[h!]
\centering
\includegraphics[width=0.6\textwidth]{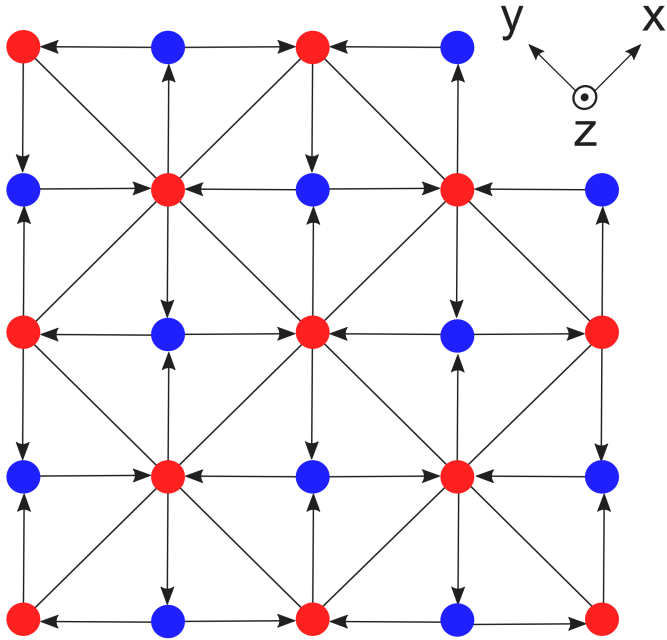}
\caption{The Union Jack lattice. Red (blue) circles represent the A (B) sublattice. The sign of the DM interaction is $\nu_{ij}=+1 (-1)$ if $i,j$ is following (against) the arrows.}%
\label{rede_UJ}
\end{figure}

We obtain the Union Jack lattice by adding alternate diagonals to the square
lattice (Figure \ref{rede_UJ}). The lattice is divided into A and B sublattices, denoted by indices $i$ and $j$, respectively. Here $\left\langle i,j\right\rangle$ means the near-neighbor (NN) spins with exchange interaction between sites A and B, and $\left\langle \left\langle i,i'\right\rangle \right\rangle $ means next-near-neighbor (NNN) interactions between sites AA. The NNN exchange
interaction carries two kinds of anisotropy: an off-plane $\lambda>1$ anisotropy which favors the alignment of the spins in the $\mathbf{\hat{z}}$ direction; and
an in-plane anisotropy in the form of different exchange constants $J_{2,ii'}$ for different in-plane directions ($J_{2,x}=J_{2}$ and $J_{2,y}=\alpha J_{2}$).

The third term of the Hamiltonian is the spin-orbit-induced Dzyaloshinskii-Moriya interaction (DMI) between sites A and B. It has become well established that DMI is the primary source of topological
magnon effects in quantum magnets. The term has the general form $\mathbf{D_{ij}} \cdot \mathbf{S}_{i} \times \mathbf{S}_{j}$ with $\mathbf{D_{ij}} = -\mathbf{D_{ji}}$. We take the interaction along the $\mathbf{\hat{z}}$ direction so that $\mathbf{D_{ij}} = D \nu_{ij} \mathbf{\hat{z}}$ with $\nu_{ij}=\pm1$ for different bond orientations (see Figure \ref{rede_UJ}). For sufficiently large $D$ (comparable to $J_{1}$) the
ground state is no longer a collinear antiferromagnet but a spin spiral. The magnetic sublattices tilt away from their antiparallel alignment, forming a net magnetic moment known as weak ferromagnetic moment. Therefore, we will consider only small $D$ values compared to $J_1$, which would preserve the Néel ground state.

The occurrence of DMI requires that the spatial inversion symmetry of the crystal field surrounding the magnetic ions be broken. Hence, the conventional DM term is absent if the crystal lattice is centrosymmetric. However, this term can be induced by an
external electric field, which has the advantage that the strength of the
field can be tuned \cite{Katsura2005}. 

When the DM term is imaginary in the reciprocal space Hamiltonian, we have an analogue of the Aharonov-Casher effect mentioned in Section 2, with $\mathbf{D}_{ij}\propto\mathbf{E} \times \mathbf{\hat{e}}_{ij}$ (where $\mathbf{\hat{e}}_{ij}$ is the unit vector connecting the two sites $i$ and $j$). In this case, the DMI contributes to the Berry curvature and is responsible for the Lorentz force acting on the propagating magnons (the DM vector acts as the vector potential or a gauge field to the spin current). That is what happens in the Union Jack lattice. On the other hand, if the DM term is real, it does not contribute to the Berry curvature. As we need an imaginary term in the Hamiltonian for a nun-null Berry curvature, it has to come from another term, such as the structure factor of the exchange interaction. This is the case of the honeycomb and brick-wall lattices (see Section 8), where we have a non-null Berry curvature even without DM interaction.

The last term in Hamiltonian (\ref{hamilt}) is a single ion easy-axis anisotropy (SIA)
which favors spin alignment along the $\mathbf{\hat{z}}$ direction. Magnetic anisotropy is crucial to overcome thermal fluctuations and stabilize the magnetic order. Since the Hamiltonian is invariant under global spin rotation about the z-axis, the $\mathbf{\hat{z}}$ component of the total spin is conserved. 

We set the spacing of each sublattice equal to 1, and take the $\mathbf{\hat{x}}$ and $\mathbf{\hat{y}}$ directions along the diagonal axes of the two sublattices since the lattice is translationally invariant along the diagonals. The Hamiltonian (\ref{hamilt}) with $A=D=0$ and $\lambda=\alpha=1$ was studied by
Collins et al. \cite{Collins2006} at zero temperature. For these values of parameters,
our calculations agree with the ones performed by them. The lattice is
frustrated, and in particular, there is a phase transition from an ordered
N\'{e}el phase to a canted ferrimagnetic phase at $\eta_{c}=J_{2}/J_{1}%
\approx0.65$ for spin $1/2$ \cite{Bishop2010,Zheng2007} and $\eta_{c}\approx0.58$ for $S=1$ \cite{Bishop2011}.
Hence, we restrict the values of the parameters to the regime where the
N\'{e}el order is preserved. The term $A>0$ favors easy-axis alignment along
the z-axis and leads to an ordered phase below the transition temperature
$T_{N}$.

As usual, we take the N\'{e}el order perpendicular to the lattice plane, i. e., spins on the A and B sublattices satisfy $\mathbf{S}_{A}=-\mathbf{S}_{B}%
=S\mathbf{\hat{z}}$ in the ground state. We use the linearized Holstein-Primakoff
representation for antiferromagnets

\begin{align}
S_{i}^{+}  &  =\sqrt{2S}a_{i}\,\,\,,\,\,S_{i}^{-}=\sqrt{2S}a_{i}^{\dagger
}\,\,,\,\,S_{i}^{z}=S-a_{i}^{\dagger}a_{i}\text{ \ \ \ \ \ \ \ }i\in
A\nonumber\\
S_{j}^{+}  &  =\sqrt{2S}b_{j}^{\dagger}\,\,\,,\,\,S_{j}^{-}=\sqrt{2S}%
b_{j}\,\,,\,\,S_{i}^{z}=-S+b_{j}^{\dagger}b_{j}\text{ \ \ \ \ \ }j\in B
\label{HP}%
\end{align}

The magnetic excitations can be described as chargeless bosonic quasiparticles (magnons) carrying a dipole momentum $\sigma g \mu_B \mathbf{\hat{e}_z}$ with $\sigma = \pm1$. Since we are considering only single-magnon excitations, the average magnon
number on each spin is much smaller than $2S$. The approximation is valid when the spin magnitude $S$ is large and/or the temperature is low enough, such that the population of thermally activated magnons at each site becomes small. However, the theory was used to fit experimental data in an $S = 1/2$ compound and remarkably gave better results than the Schwinger boson representation, even above the transition temperature \cite{Hirschberger2015}.

We perform a Bogoliubov transformation from bosons $(a_{i},b_{i})$ to
$(\alpha_{i},\beta_{i})$ (see Section 4). Using Eqs. (\ref{HP}) and
(\ref{transform}) we can show that

\begin{equation}
S^{z}=\sum_{i\in A}\sum_{j\in B}\left(  S_{i}^{z}+S_{j}^{z}\right)  =\sum
_{k}\left(  -\alpha_{k}^{\dagger}\alpha_{k}+\beta_{k}^{\dagger}\beta
_{k}\right)
\end{equation}

Hence $\left\langle 0\right\vert \alpha_{k}S^{z}\alpha_{k}^{\dagger}\left\vert
0\right\rangle =-1$ and $\left\langle 0\right\vert \beta_{k}S^{z}\beta
_{k}^{\dagger}\left\vert 0\right\rangle =+1$ (where $\left\vert 0\right\rangle
$ is the magnon vacuum), showing that $\alpha$ magnons carry $-1$ spin angular
momentum and $\beta$ magnons carry $+1$ spin angular momentum along the $\mathbf{\hat{z}}$
direction \cite{Cheng2016}. Momentum and spin conservations are fulfilled by a
combination of $\alpha$ and $\beta$ magnons, which carry opposite spins and
momenta ($\mathbf{k}$ and $-\mathbf{k}$). Note that the magnon operators $\alpha$
and $\beta$ are a mixture of $a$ and $b$ operators (which are not magnon
operators), and therefore are not associated to sublattices A and B.

Taking Eq. (\ref{HP}) into the Hamiltonian (\ref{hamilt}) and Fourier
transforming, we obtain (after discarding the zero-point energy and higher
order terms) the following quadratic Hamiltonian:

\begin{equation}
H=H_{1}+H_{2}+H_{DM}+H_{SIA}%
\end{equation}

where

\begin{align}
H_{1}  &  =2J_{1}S\sum_{k}\left[  a_{k}^{\dagger}a_{k}+a_{k}a_{k}^{\dagger
}+b_{k}^{\dagger}b_{k}+b_{k}b_{k}^{\dagger}+\gamma_{k}\left(  a_{k}%
b_{-k}+b_{-k}a_{k}+a_{k}^{\dagger}b_{-k}^{\dagger}+b_{-k}^{\dagger}%
a_{k}^{\dagger}\right)  \right] \\
H_{2}  &  =SJ_{2}\sum_{k}\left[  \left(  2\eta_{k}-\lambda\left(
\alpha+1\right)  \right)  \left(  a_{k}^{\dagger}a_{k}+a_{k}a_{k}^{\dagger
}\right)  \right] \\
H_{DM}  &  =2iSD\sum_{k}m_{k}\left[  \left(  a_{k}b_{-k}+b_{-k}a_{k}\right)
-\left(  a_{k}^{\dagger}b_{-k}^{\dagger}+b_{-k}^{\dagger}a_{k}^{\dagger
}\right)  \right] \\
H_{SIA}  &  =\frac{1}{2}A\left(  2S-1\right)  \sum_{k}\left(  a_{k}^{\dagger
}a_{k}+a_{k}a_{k}^{\dagger}+b_{k}^{\dagger}b_{k}+b_{k}b_{k}^{\dagger}\right)
\label{SIA}%
\end{align}

with the structure factors defined as

\begin{equation}
\gamma_{k}=\cos\frac{k_{x}}{2}\cos\frac{k_{y}}{2},\text{ \ \ \ \ \ \ \ \ \ }%
\eta_{k}=\frac{1}{2}\left(  \cos k_{x}+\alpha\cos k_{y}\right)  \text{,
\ \ \ \ \ \ \ }m_{k}=-\sin\frac{k_{x}}{2}\sin\frac{k_{y}}{2} \label{UJ_structure}
\end{equation}

For the appearance of the term $\left(  2S-1\right)  $ in Eq. (\ref{SIA}) see
Eq. (\ref{SIA_2S1}). We see that for $S=1/2$, we have $H_{SIA}=0$ and the single-ion anisotropy is not effective.

We show the calculation explicitly for $m_{k}$ considering the factor $\nu_{ij}$ as $+1$ in the up/down direction and
$-1$ in the left/right direction, departing from an A site (see Figure (\ref{rede_UJ}):

\begin{equation}
m_{k}=\frac{1}{4}\left(  e^{i\left(  k_{x}+k_{y}\right)  /2}-e^{i\left(
-k_{x}+k_{y}\right)  /2}-e^{i\left(  k_{x}-k_{y}\right)  /2}+e^{-i\left(
k_{x}+k_{y}\right)  /2}\right)  =-\sin\frac{k_{x}}{2}\sin\frac{k_{y}}{2}%
\end{equation}

As mentioned above, the spin wave formalism, although widely used in the
literature, is justified only for large spin values and low
temperatures. Note, however, that our calculations are valid for any spin value.

To present a general discussion of an antiferromagnet, we write the Hamiltonian
(\ref{hamilt}) as

\begin{align}
H  &  =\sum_{k}\left(  M_{11}a_{k}^{\dagger}a_{k}+M_{12}a_{-k}^{\dagger}%
b_{k}^{\dagger}+M_{21}b_{-k}a_{k}+M_{22}b_{k}b_{k}^{\dagger}+M_{33}a_{k}%
a_{k}^{\dagger}+\right. \nonumber\\
&  \text{ \ \ \ \ \ \ \ \ \ \ }+\left.  M_{34}a_{-k}b_{k}+M_{43}%
b_{-k}^{\dagger}a_{k}^{\dagger}+M_{44}b_{k}^{\dagger}b_{k}\right)
\end{align}

This can be written as

\begin{equation}
H=\sum_{k}\psi_{k}^{\dagger}H_{k}\psi_{k}%
\end{equation}

Here, for convenience, we use a different convention that that used in Eq. (\ref{basis_old}), and write $\psi_{k}^{\dagger}=\left(
\begin{array}
[c]{cccc}%
a_{k}^{\dagger} & b_{-k} & a_{-k} & b_{k}^{\dagger}%
\end{array}
\right)$. \textcolor{black}{This k-space basis is similar to the one used in Ref. \cite{Zyuzin2016}. As a result, the matrix $H_{k}$ splits in two blocks:}

\begin{equation}
H_{k}=\left(
\begin{array}
[c]{cc}%
M_{k} & 0\\
0 & M_{-k}^{\ast}%
\end{array}
\right)  \text{, \ \ where }M_{k}=\left(
\begin{array}
[c]{cc}%
r_{1k} &   f_{k}  ^{\ast}\\
f_{k} & r_{2k}%
\end{array}
\right)  \text{\ } \label{hamilt_matrix}%
\end{equation}

We can notice that the $\mathbf{k}$ dependence comes only from the structure
factors $\gamma_{k},$ $\eta_{k}$ and $m_{k}$. For the Union Jack lattice as studied here, these factors are even functions of $\mathbf{k}$. Hence, $M_{k}=M_{-k}$ and we can
ignore the sign of $\mathbf{k}$ in the Hamiltonian parameters. We identify

\begin{equation}
M_{11}=r_{1k,}\text{ \ \ }M_{22}=r_{2k},\text{ \ \ }M_{21}=f_k=h_{xk}+ih_{yk},\text{
\ }M_{12}=f_k^{\ast}=h_{xk}-ih_{yk} \label{M}
\end{equation}

Specifically for the Union Jack lattice, we have:

\begin{align}
r_{1k}  &  =S\left[  2J_{1}+J_{2}\left(  2\eta_{k}-\lambda\left(
\alpha+1\right)  \right)  \right]  +\frac{1}{2}A\left(  2S-1\right)
\nonumber\\
r_{2k}  &  =2SJ_{1}+\frac{1}{2}A\left(  2S-1\right) \nonumber\\
h_{xk}  &  =2SJ_{1}\gamma_{k}\nonumber\\
h_{yk}  &  =2SDm_{k} \label{parameters}%
\end{align}

A Hamiltonian with particle-hole symmetry is known as the
Bogoliubov-de Gennes (BdG) Hamiltonian and is written as

\begin{equation}
H_{BdG}=\left(
\begin{array}
[c]{cc}%
\Xi_{k} & \Lambda_{k}\\
\Lambda_{-k}^{\ast} & \Xi_{-k}^{\ast}%
\end{array}
\right)
\end{equation}

where $\Xi_{k}$ and $\Lambda_{k}$ are $2\times2$ matrices. The Hamiltonian for the AF
honeycomb \cite{Owerre8,Cheng2016,Zyuzin2016,Zhang2022} and square \cite{Kawano2019} lattices can be written this way. For the Union Jack lattice, we see that $\Xi_{k}=M_{k}$ and
$\Lambda_{k}=0$.

From Eq. (\ref{hamilt_matrix}), and noting the specific order of the operators in $\psi
_{k}^{\dagger}$, we have

\begin{equation}
\eta=diag\left(  1,-1,-1,1\right)  =\left(
\begin{array}
[c]{cc}%
\sigma_{z} & 0\\
0 & -\sigma_{z}%
\end{array}
\right)
\end{equation}

The advantage of working with a block-diagonal Hamiltonian is evident when we
write $K_{k}$ as

\begin{equation}
K_{k}=\eta H_{k}=\left(
\begin{array}
[c]{cc}%
\sigma_{z}M_{k} & 0\\
0 & -\sigma_{z}M_{-k}^{\ast}%
\end{array}
\right)
\end{equation}

We can, then, diagonalize the sectors of $K_{k}$ separately. The first sector,
which we call $\alpha$-sector, has eigenvectors

\begin{equation}
\varphi_{\alpha}^{\left(  +\right)  }=\left(
\begin{array}
[c]{c}%
u_k^{\ast}\\
-v_k^{\ast}%
\end{array}
\right)  ,\text{ \ \ \ \ \ \ }\varphi_{\alpha}^{\left(  -\right)  }=\left(
\begin{array}
[c]{c}%
-v_k\\
u_k
\end{array}
\right)
\end{equation}

with corresponding eigenvalues

{\color{black}
\[
\omega_{\alpha}^{\left(  +\right)}(\mathbf{k})=w(\mathbf{k})+\Delta(\mathbf{k}),\text{ \ \ }\omega_{\alpha
}^{\left(  -\right)  }(\mathbf{k})=-w(\mathbf{k})+\Delta(\mathbf{k})
\]
}

The parameters $u$ and $v$ are defined as

\begin{equation}
u=\frac{f}{\left\vert f\right\vert }\left(  \frac{r+w}{2w}\right)
^{1/2},\text{ \ \ \ \ \ \ }v=v^{\ast}=\left(  \frac{r-w}{2w}\right)  ^{1/2}%
\end{equation}

with%

\begin{equation}
r=\frac{r_{1}+r_{2}}{2},\text{ \ \ }\Delta=\frac{r_{1}-r_{2}}{2},\text{
\ \ }w=\sqrt{r^{2}-\left\vert f\right\vert ^{2}}\text{\ \ } \label{randDelta}%
\end{equation}

The eigenvectors of the $\alpha$-sector can also be written as \cite{Owerre8}

\begin{equation}
\varphi_{\alpha}^{\left(  +\right)  }=\left(
\begin{array}
[c]{c}%
e^{-i\phi}\cosh\frac{\theta}{2}\\
-\sinh\frac{\theta}{2}%
\end{array}
\right)  ,\text{ \ \ \ \ \ \ }\varphi_{\alpha}^{\left(  -\right)  }=\left(
\begin{array}
[c]{c}%
-\sinh\frac{\theta}{2}\\
e^{i\phi}\cosh\frac{\theta}{2}%
\end{array}
\right)
\end{equation}

where the new parameters $\phi$ and $\theta$ are defined as

\begin{equation}
\tan\phi=\frac{h_{y}}{h_{x}},\text{ \ \ }\cosh\theta=\frac{r}{w} \label{thetaphi}%
\end{equation}

This form is particularly useful to obtain an expression for the Berry
curvature, as it will be shown later. We note that

\begin{equation}
e^{i\phi}=\frac{f}{\left\vert f\right\vert }\text{, \ \ }\cosh\frac{\theta}%
{2}=\left(  \frac{r+w}{2w}\right)  ^{1/2}\text{, \ \ }\sinh\frac{\theta}%
{2}=\left(  \frac{r-w}{2w}\right)  ^{1/2}%
\end{equation}

There is a degree of freedom in choosing the eigenvectors, which are related by $U(1)$
gauge transformations. That is, the eingenstates can be written in two ways related by a phase $e^{i \phi}$. However, the gauge does not affect the Berry curvature.

The eigenvectors of the $\beta$-sector (second block of $H_{k}$) are the
complex conjugate of the $\alpha$-sector \textcolor{black}{with a substitution $\mathbf{k} \rightarrow -\mathbf{k}$. The eigenvalues are:}

{\color{black}
\[
\omega_{\beta}^{\left(  +\right)}(\mathbf{k})=w(\mathbf{-k})-\Delta(\mathbf{-k}),\text{ \ \ }\omega_{\beta
}^{\left(  -\right)  }(\mathbf{k})=-w(\mathbf{-k})-\Delta(\mathbf{-k}) \nonumber
\]
}

As we saw before, a quantum of $\alpha$-magnon carries spin $-1$
and a quantum of $\beta$-magnon a spin $+1$. The transformation doubles the
Hilbert space, so the eigenvalues of $K_{k}$ show up in pairs \textcolor{black}{$\pm\omega(\pm \mathbf{k})$}.
For bosons, only positive energy states are physical, and we keep only the
positive branches $\varphi_{\alpha}^{\left(  +\right)  }$ and $\varphi_{\beta
}^{\left(  +\right)  }$ \cite{Samajdar2019}. \textcolor{black}{Nevertheless, when using the Kubo formula we must perform the calculations in the full particle-hole space. Then, using the particle-hole symmetry we can express our result only in terms of the positive energy states \cite{Li2020}. The matrix $T_{k}$,
which diagonalizes $K_{k}$ and carries all four eigenvectors is}:

{\color{black}

\begin{align}
T_{k}  &  =\left(
\begin{array}
[c]{cccc}%
u^{\ast}_k & -v_k & 0 & 0\\
-v^{\ast}_k & u_k & 0 & 0\\
0 & 0 & u_{-k} & -v_{-k}^{\ast}\\
0 & 0 & -v_{-k} & u_{-k}^{\ast}%
\end{array}
\right) \label{Tk}\\
&  =\left(
\begin{array}
[c]{cccc}%
e^{-i\phi_k}\cosh\frac{\theta_k}{2} & -\sinh\frac{\theta_k}{2} & 0 & 0\\
-\sinh\frac{\theta_k}{2} & e^{i\phi_k}\cosh\frac{\theta_k}{2} & 0 & 0\\
0 & 0 & e^{i\phi_{-k}}\cosh\frac{\theta_{-k}}{2} & -\sinh\frac{\theta_{-k}}{2}\\
0 & 0 & -\sinh\frac{\theta_{-k}}{2} & e^{-i\phi_{-k}}\cosh\frac{\theta_{-k}}{2}%
\end{array}
\right) \nonumber\\
&  =\left(  \varphi_{\alpha}^{\left(  +\right)  },\varphi_{\alpha}^{\left(
-\right)  },\varphi_{\beta}^{\left(  -\right)  },\varphi_{\beta}^{\left(
+\right)  }\right) \nonumber
\end{align}

and the diagonalized $K_{k}$, which carries the eigenvalues, is:

\begin{equation}
E_{k}\equiv T_{k}^{-1}K_{k}T_{k}=\left(
\begin{array}
[c]{cccc}%
\omega_{\alpha}^{\left(  +\right)  }(\mathbf{k}) & 0 & 0 & 0\\
0 & \omega_{\alpha}^{\left(  -\right)  }(\mathbf{k}) & 0 & 0\\
0 & 0 & \omega_{\beta}^{\left(  -\right)  }(\mathbf{k}) & 0\\
0 & 0 & 0 & \omega_{\beta}^{\left(  +\right)  }(\mathbf{k})%
\end{array}
\right)   \nonumber
\end{equation}

The positive energy spectrum is given by:

\begin{align}
E_\alpha^{\left(  +\right)  }/ \hbar = \omega_{\alpha}^{\left(  +\right)  }(\mathbf{k})  &  =w(\mathbf{k}) + \Delta(\mathbf{k}) \nonumber\\
E_\beta^{\left(  +\right)  }/ \hbar = \omega_{\beta}^{\left(  +\right)  }(\mathbf{k})  &  =w(-\mathbf{k})-\Delta(-\mathbf{k}) \label{spectrum}%
\end{align}
}

Here, the particle states correspond to $T_k$'s first and fourth columns. That comes from the specific order of the operators chosen in the
vector $\psi_{k}^{\dagger}=\left(
\begin{array}
[c]{cccc}%
a_{k}^{\dagger} & b_{-k} & a_{-k} & b_{k}^{\dagger}%
\end{array}
\right)  $. The second and third columns represent the hole states with a negative energy spectrum (non-physical). We analyze the band structure in the following.

\textcolor{black}{For the Union Jack lattice we can ignore the minus sign in the argument of the functions in the $\beta$-sector, as all the structure factors are even in $\mathbf{k}$ (See Eq. (\ref{UJ_structure}))}. With definitions (\ref{parameters}) and (\ref{randDelta}), we see that

\begin{equation}
\Delta(\mathbf{k})=\frac{r_{1}-r_{2}}{2}=\frac{1}{2}SJ_{2}\left[  2\eta_{k}-\lambda\left(
\alpha+1\right)  \right]
\end{equation}

\begin{figure}[h!]
\centering
\includegraphics[width=0.6\textwidth]{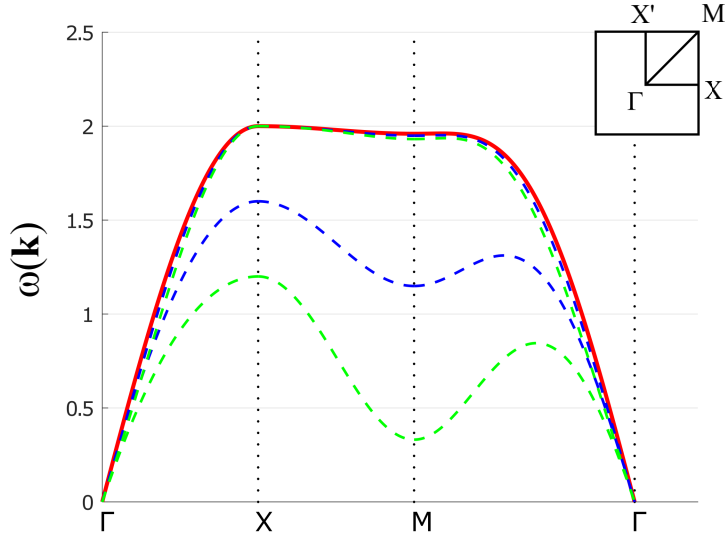}\caption{Energy bands of the AF Union Jack lattice for
three values of $J_{2}/J_{1}$: 0 (square lattice, solid red, both bands are
totally degenerate), 0.2 (dashed blue) and 0.4 (dashed green). Other
parameters are $S=J_{1}=\lambda=\alpha=1$, $A=0$, $D=0.2$. Both bands
have a null gap.} \label{bands_J2}
\end{figure}

If $J_{2}=0$ the term $\Delta$ vanishes, and the magnon bands are degenerate. The system is reduced to the AF square lattice \cite{Nakata2017}. Higher values of
$J_{2}$ lower the band of the $\alpha$-mode, while the $\beta$-mode band
remains almost unchanged (Figure \ref{bands_J2}).

\begin{figure}[h!]
\centering
\includegraphics[width=0.6\textwidth]{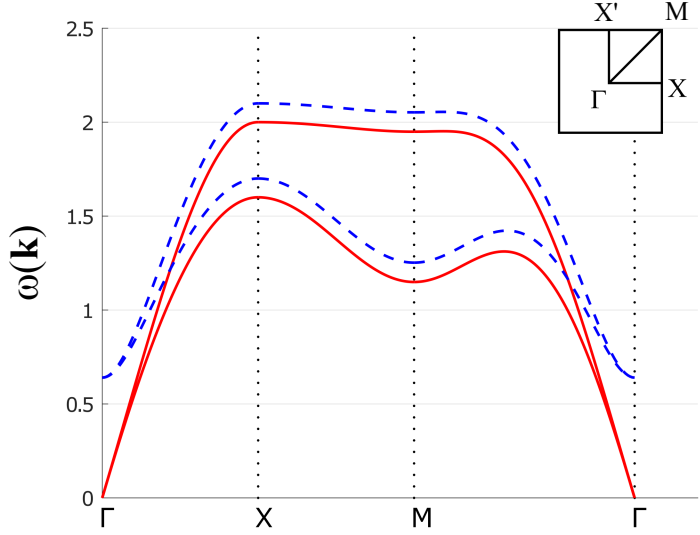}\caption{Energy bands of
the AF Union Jack lattice with $A=0$ (solid red) and $A=0.2$ (dashed blue). Other parameters
are $S=J_{1}=\lambda=\alpha=1$, $J_{2}=D=0.2$. The SIA opens the gap, but
bands are still degenerate at point $\Gamma$.} \label{bands_A}
\end{figure}

\begin{figure}[h!]
\centering
\includegraphics[width=0.6\textwidth]{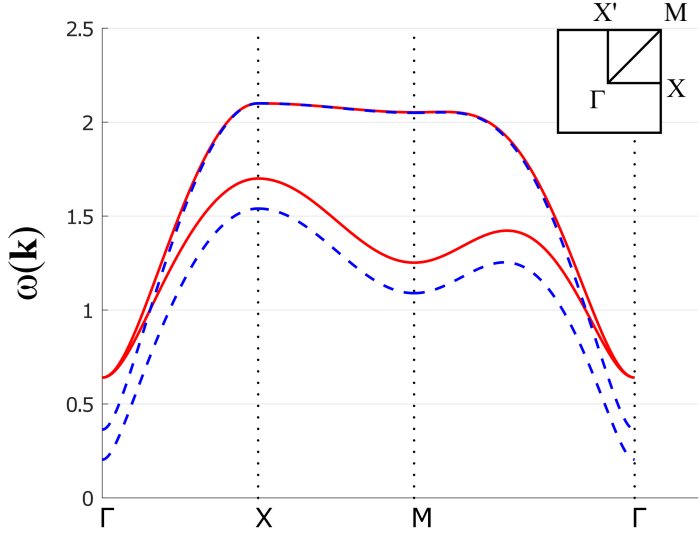}\caption{Dispersion curves of
the AF Union Jack lattice for $\lambda=1$ (solid red) and $\lambda=1.4$ (dashed blue). Other
parameters are $S=J_{1}=\alpha=1$, $J_{2}=A=D=0.2$. The off-plane exchange
anisotropy $\lambda$ splits and lowers the bands at point $\Gamma$.} \label{bands_lambda}
\end{figure}

\begin{figure}[h!]
\centering
\includegraphics[width=0.6\textwidth]{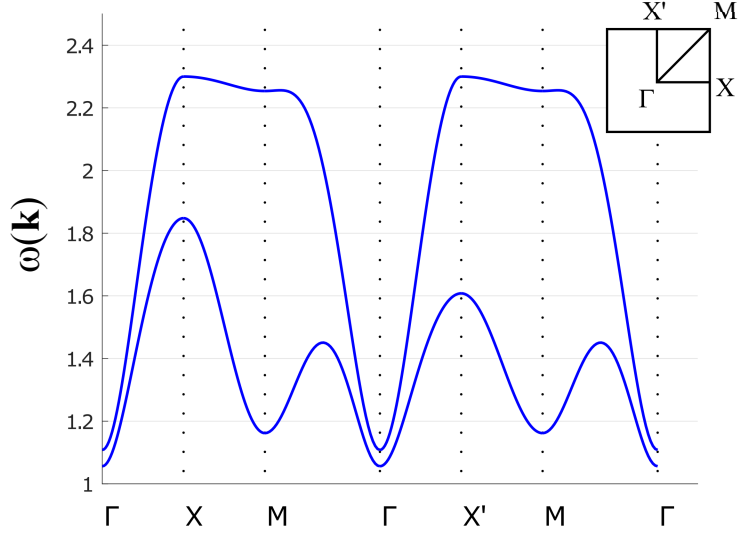}\caption{The effect of the
in-plane exchange anisotropy $\alpha$ on the AF Union Jack lattice is to create an energetic inequivalence
between points axes $k_x$ and $k_y$. The parameters are
$S=J_{1}=1$, $J_{2}=D=0.2$, $A=0.6$, $\lambda=1.1$, $\alpha=1.6$.} \label{bands_alpha}
\end{figure}

\begin{figure}[h!]
\centering
\includegraphics[width=0.6\textwidth]{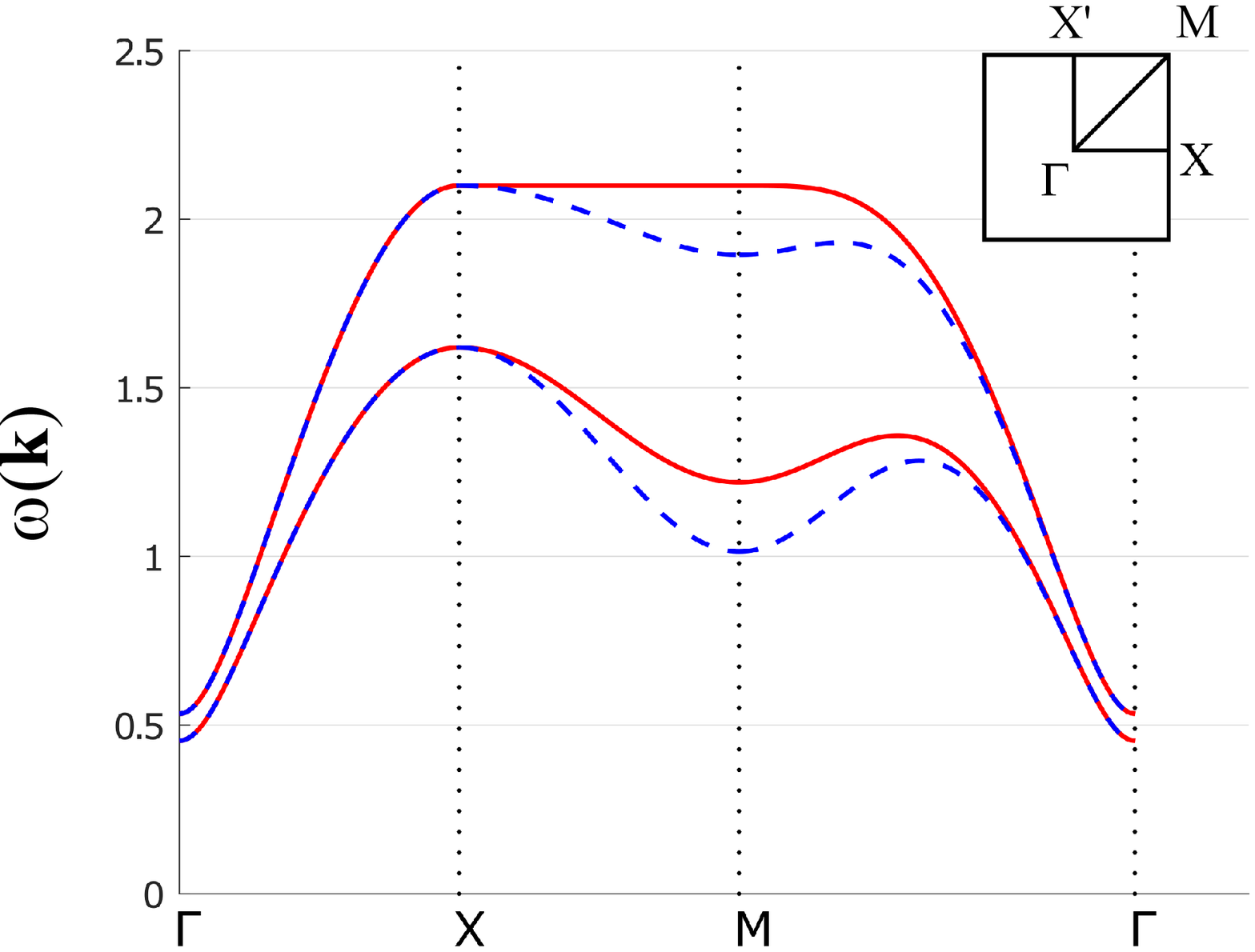} \caption{Effect of the DM
interaction on the dispersion curves of the AF Union Jack lattice. The solid red line is the $D=0$ case, while
the blue dashed line is the $D=0.4$ case. Other parameters are $S=J_{1}%
=\alpha=1$, $J_{2}=A=0.2$, $\lambda=1.2$.} \label{bands_D}
\end{figure}

For small enough values of $A/J_{1},$
$J_{2}/J_{1}$ and $D/J_{1}$ both bands have a minimum at $k_{x}=k_{y}=0$
($\Gamma$ point). This minimum defines the bands' gaps:

\begin{equation}
\omega_{gap}^{\alpha,\beta}=w_{0} \pm\Delta_{0}%
\end{equation}

with

\begin{equation}
\Delta_{0}=\frac{1}{2}SJ_{2}\left(  \alpha+1\right)  \left(  1-\lambda\right)
\end{equation}

We see that for $\lambda=1$ we have $\Delta_{0}=0,$ and the gap has the same
value $w_{0}$ for both bands, meaning there is a degeneracy at point
$\Gamma$. In this picture, we have:

\begin{equation}
\left.  \omega_{gap}\right\vert _{\lambda=1}=\frac{1}{2}\sqrt{A\left(
2S-1\right)  }\sqrt{8SJ_{1}+A\left(  2S-1\right)  }%
\end{equation}

The gap vanishes only for $A=0$ or $S=1/2$. In other words, the $SIA$ term opens the gap at
the $\Gamma$ point (Figure \ref{bands_A}), while the off-plane exchange anisotropy
$\lambda$ makes the gap different for each magnon, splitting the bands at this
point (Figure \ref{bands_lambda}). For any $T > 0$, a null gap $\omega(\mathbf{k_0})=0$ makes the magnon population explode at point $\mathbf{k_0}$, resulting in divergent Berry curvature. Hence, we need $A\neq0$ and $S\neq1/2$ for well-behaved Berry curvature and transverse transport coefficients,. The non-null gap stabilizes the ground state as it becomes energetically isolated from the rest of the spectrum.

The effect of the in-plane $\alpha$ anisotropy is to create an energetic
inequivalence between points $X$ and $X^{\prime}$ of the Brillouin zone
(Figure \ref{bands_alpha}). This energetic imbalance enables non-null transverse transport. We also note that the DMI itself doesn't open a gap, but changes
the character of the dispersion by lowering the energy at the point $M$
(Figure \ref{bands_D}).

\begin{figure}[h!]
\centering
\includegraphics[width=0.6\textwidth]{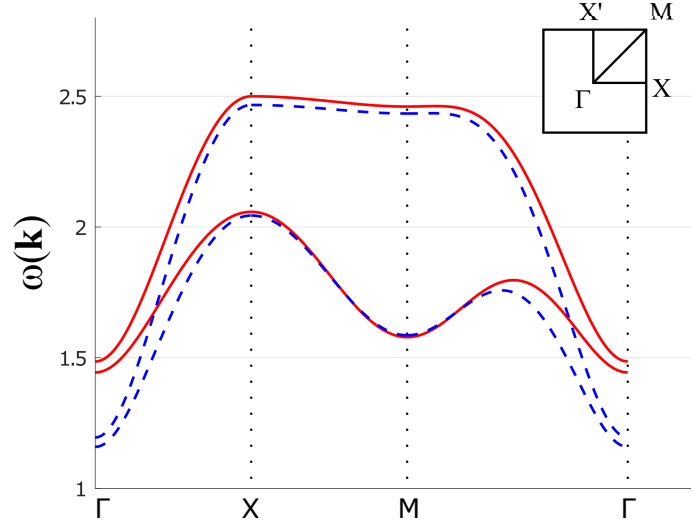}\caption{Linear spin wave dispersion relation (LSW,
solid red line) compared to the modified spin wave case (MSW, dashed blue line) \textcolor{black}{in} the Union Jack lattice.
The temperature is zero, parameters are $S=J_{1}=A=1$, $J_{2}=D=0.2$,
$\lambda=\alpha=1.1$. The MSW lowers the gap even at zero temperature.} \label{bands_MSW}
\end{figure}

\begin{figure}[h!]
\centering
\includegraphics[width=0.6\textwidth]{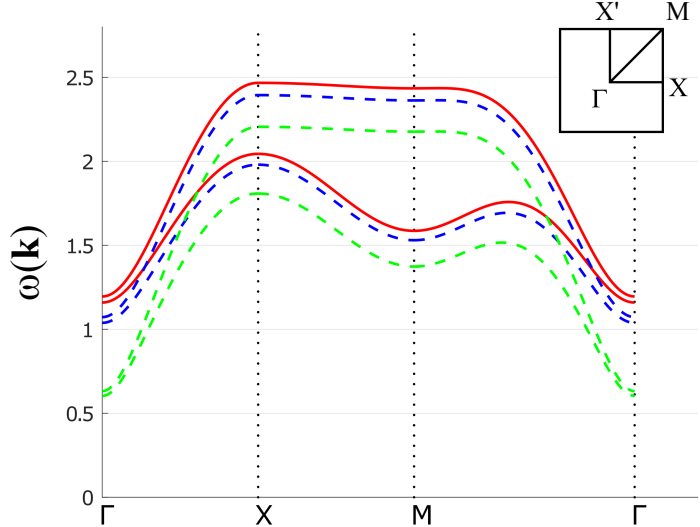} \caption{MSW dispersion for
$T=0$ (solid red), $T=0.4$ (dashed blue) and $T=0.455$ (dashed green) in the Union Jack lattice. The gap
shrinks abruptly as we approach the transition temperature $T_{N}
\approx0.456$. The parameters are $S=J_{1}=A=1$, $J_{2}=D=0.2$, $\lambda
=\alpha=1.1$.} \label{bands_MSW_T}
\end{figure}

\begin{figure}[h!]
\centering
\includegraphics[width=0.6\textwidth]{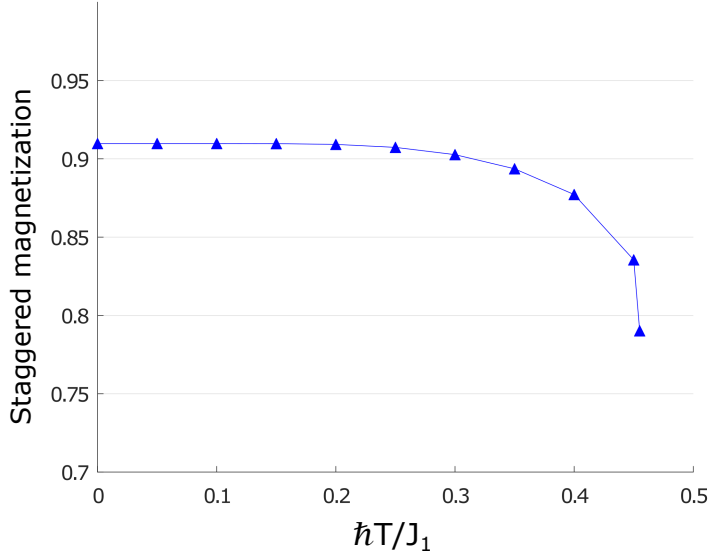}\caption{Staggered
magnetization for MSW on the Union Jack lattice. The parameters are $S=J_{1}=A=1$,$J_{2}=D=0.2$,
$\lambda=\alpha=1.1$. The
Néel temperature was estimated in $T_{N} \approx0.456$.} \label{magn}
\end{figure}

We also studied the effect of the anharmonic contributions (magnon-magnon interactions) in a
mean-field picture, where the Hamiltonian is renormalized by
temperature-dependent parameters. In this modified spin wave (MSW) approach, even at zero temperature, the anharmonic contributions affect the energy bands, lowering the bands' gap (Figure \ref{bands_MSW}). As the temperature rises, the bands remain virtually unchanged until we get closer to the Néel temperature $T_{N}$, when the
energy drops abruptly (Figure \ref{bands_MSW_T}). At $T_{N}$ the Néel order is unstable, and
the staggered magnetization vanishes, signaling a phase
transition (Figure \ref{magn}). The transition temperature $T_{N}$ tends to be higher
for high values of $A$ or $S$. A complete description and the analytical results of the MSW approach can be found in Appendix A.

\section{Symmetries}

Crystal structures can be classified by the symmetries they present. An ideal crystal is a periodically repeating pattern. Since all lattice points of a periodic lattice are equivalent, every point has the same neighborhood as other points. So the lattice itself can be characterized by listing the symmetry operations that keep each of them fixed \cite{Bonbien2021}. The point symmetries are rotations $c_n$ (due to periodicity, rotations are restricted to be $360^{\circ}/n$, where $n = 1,2,3,4,6$), reflections, space inversions, and combinations of them. Space inversion maps $\mathbf{r}\rightarrow -\mathbf{r}$ and $\mathbf{k}\rightarrow -\mathbf{k}$. When we have magnetic ordering, the crystallographic symmetries can be extended to describe magnetic crystals. Spatial inversion symmetry implies $\omega \left( \mathbf{k} \right) = \omega \left( \mathbf{-k} \right)$.

The time-reversal operation $\mathcal{T}$ inverts all spins, changes $\mathbf{k}$ into $-\mathbf{k}$ and takes the complex conjugate. Note  that a Hamiltonian can be invariant under TRS, but the symmetry can be spontaneously broken in the ground state. In an antiferromagnet with Néel order TRS is broken, since flipping all the spins would swap the A and B sublattices, changing the sign of the staggered magnetization. Another symmetry is broken, namely, the symmetry under lattice translation $T_\mathbf{a}$, where $\mathbf{a}$ is a vector connecting two sublattices in the case of a bipartite antiferromagnet. In general, time-reversal symmetry can be effectively restored if another symmetry $\mathcal{S}$ of the Hamiltonian, combined with the TRS, is a good symmetry of the system. For instance, besides translation, the Hamiltonian can also remain invariant under the combined time-reversal and spin rotation ($\mathcal{S}=c_{x}$) by $180^\circ$ around the x-axis (or any other in-plane axis). These combined symmetries operations $\mathcal{T}\mathcal{S}$ are sometimes called effective time-reversal symmetry (ETRS). 

To understand the symmetry properties of the ground state of the Heisenberg antiferromagnet, we follow Cheng et al. \cite{Cheng2016} and start with the Néel ground state. We write

\begin{align}
H&=J_1 \sum_{\left\langle A,B \right\rangle} \mathbf{S}_A \cdot \mathbf{S}_B + J_2 \sum_{\left\langle \left\langle A,A' \right\rangle\right\rangle} \mathbf{S}_A \cdot \mathbf{S}_{A'} + \mathcal{K} \sum_{A,B} \left[ \left( S_A^z \right) ^2 +\left( S_B^z \right) ^2 \right] \nonumber \\
&= H_{1} + H_{2} + H_{SIA}
\label{sym_hamiltonian}
\end{align}

where $J_1$ is an exchange interaction between NN sites of different sublattices, and $J_2$ is between NNN of the same sublattice A (it could as well be sublattice B). Writing $\mathbf{S}_A= \delta \mathbf{S}_A + \mathbf{\hat{z}}$ and $\mathbf{S}_B= \delta \mathbf{S}_B - \mathbf{\hat{z}}$, we have

\begin{align}
\mathbf{S}_A \cdot \mathbf{S}_B &= \left( \delta \mathbf{S}_A + \mathbf{\hat{z}} \right) \cdot  \left(  \delta \mathbf{S}_B - \mathbf{\hat{z}} \right) = \delta \mathbf{S}_A \cdot \delta \mathbf{S}_B - \delta S_A^z +\delta S_B^z -1 \nonumber \\
\mathbf{S}_A \cdot \mathbf{S}_{A'} &= \left( \delta \mathbf{S}_A + \mathbf{\hat{z}} \right) \cdot  \left(  \delta \mathbf{S}{A'} + \mathbf{\hat{z}} \right) = \delta \mathbf{S}_A \cdot \delta \mathbf{S}_{A'} + \delta S_A^z +\delta S_{A'}^z +1 \nonumber \\
\left( S_A^z \right) ^2 +  \left( S_B^z \right) ^2 & = \left( \delta S_A^z \right) ^2 +  \left( \delta S_B^z \right) ^2 + 2 \left( 1+ \delta S_A^z -\delta S_B^z \right) \label{displace}
\end{align}

As we are interested in the symmetry properties of magnons, all symmetry operations act only on the perturbations $\mathbf{\delta S}_{A,B}$, leaving the Néel ground state unchanged \cite{Cheng2016}. The time-reversal operation $\mathcal{T}$ changes the sign of all components of $\mathbf{\delta S}$. The spin rotation $c_x$ (by $180^\circ$) only changes the sign of $\delta S^{y}$ and $\delta S^{z}$, leaving $\delta S^{x}$ unchanged. Hence, the combined symmetry $\mathcal{T}c_x$ have the effect of changing the sign of $\delta S^{x}$. From Eq. (\ref{displace}), we can see that $H_{1} + H_{2} + H_{SIA}$ is invariant under the $\mathcal{T}c_x$. This is the ETRS of the system.

Let us now consider the DM term

\begin{equation}
H_{DM}=D\sum_{i,j} \nu _{ij} \left(  S_{i}^{x}S_{j}^{y}-S_{i}^{y}S_{j}^{x}\right)
\end{equation}

Making the same substitution as before, we obtain

\begin{equation}
H_{DM}=D\sum_{ i,j } \nu _{ij} \left(  \delta S_i^{x} \delta S_j^{y}- \delta S_i^{y} \delta S_j^{x}\right)
\end{equation}

As $\mathcal{T}c_x$ changes the sign of $\delta S^{x}$, the DMI breaks the ETRS of the Hamiltonian. It is indifferent if the interaction is between sites of the same or different sublattices because the operation acts similarly in both $\mathbf{\delta S}_{A}$ and $\mathbf{\delta S}_{B}$.

The developments above are made under the assumption of a Néel ground state, but it is known that the DMI can bend the spins, generating a canted ground state. Even in a canted collinear antiferromagnet, the canting explicitly breaks time-reversal symmetry, whereas the Dzyaloshinskii-Moriya interaction provides the necessary Lorentz force on the magnons.

\section{Berry curvature and transverse transport}

The Berry curvature of a system described by a Hamiltonian like Eq. (\ref{hamilt_matrix}) can be found analytically from the Hamiltonian parameters. From Eqs. (\ref{berry_c}) and (\ref{Tk}), we can show that (see Appendix B):

{\color{black}
\begin{align}
\Omega^{\alpha}(\mathbf{k})&=-\frac{1}{2}\sinh\theta_k\left(
\frac{\partial\phi_k}{\partial k_{x}}\frac{\partial\theta_k}{\partial k_{y}}%
-\frac{\partial\phi_k}{\partial k_{y}}\frac{\partial\theta_k}{\partial k_{x}%
}\right) \nonumber \\
\Omega^{\beta}(\mathbf{k})&=-\frac{1}{2}\sinh\theta_{-k}\left(
\frac{\partial\phi_{-k}}{\partial k_{x}}\frac{\partial\theta_{-k}}{\partial k_{y}}%
-\frac{\partial\phi_{-k}}{\partial k_{y}}\frac{\partial\theta_{-k}}{\partial k_{x}%
}\right)
\end{align}
}

where $\theta_k$ and $\phi_k$ are defined in Eq. (\ref{thetaphi}). As a general result, we know that the symmetries of the Hamiltonian determine some properties of the Berry curvature. Effective time-reversal symmetry implies $\Omega\left(\mathbf{k}\right) = -\Omega \left( -\mathbf{k} \right)$ (odd function), and inversion symmetry implies $\Omega\left(\mathbf{k}\right) = \Omega \left( -\mathbf{k} \right)$ (even function). If both symmetries are present, we have $\Omega\left(\mathbf{k}\right)=0$.
 
The Berry curvature is directly related to the transversal transport effects. It acts as an ``artificial magnetic field” in momentum space, generating Hall-like transport. In the linear response theory, a magnetic field gradient can generate a spin current in the transverse direction, given by \cite{Han2017}:

\begin{equation}
j_y^{S,B} = \sigma_{xy} \left( -\partial_xB \right)
\end{equation}

That is the spin Hall effect of magnons, and we call $\sigma_{xy}$ the spin Hall conductivity. Also, the presence of a thermal gradient generates spin and heat currents, given by:

\begin{align}
j_y^{S,T} &= \alpha_{xy} \left( -\partial_xT \right) \\
j_y^{Q,T} &= \kappa_{xy} \left( -\partial_xT \right)
\end{align}

These are the spin Nernst effect of magnons and the thermal Hall effect of magnons, respectively. The coefficient $\alpha_{xy}$ is called the spin Nernst coefficient, and $\kappa_{xy}$, the thermal Hall conductivity.
The transport coefficients for a two-band antiferromagnet in the Néel state are given by \cite{Nakata2021,Han2017,Matsumoto2014,Kondo2022} (for the signs between the functions in the integrand, see \cite{Nakata2021}):

{\color{black}
\begin{align}
\sigma_{xy}&=-\frac{1}{\hbar} \int\limits_{BZ} \frac{d^2k}{\left(  2\pi\right)  ^{2}} \, \left[ n_{k}^{\alpha}\,\Omega_{k}^{\alpha}  +n_{k}^{\beta}\,\Omega_{k}^{\beta} \right]  \label{cond}\\
\alpha_{xy}  &  =-\frac{k_{B}}{\hbar} 
 \int\limits_{BZ}\frac{d^2k}{\left(  2\pi\right)  ^{2}} \text{ } \left[ c_{1}\left(  n_{k}^{\alpha
}\right)\,\Omega_{k}^\alpha - c_{1} ( n_{k}^{\beta
})\,\Omega_{k}^\beta \right]  \label{alpha}\\
\kappa_{xy}  &  =-\frac{k_{B}^{2}T}{\hbar} \int\limits_{BZ}\frac{d^2k}{\left(  2\pi\right)  ^{2}} \text{ } \left[ c_{2}\left(  n_{k}^{\alpha}\right)\,\Omega_{k}^\alpha + c_{2} ( n_{k}^{\beta
})\,\Omega_{k}^\beta \right]  \label{kappa}%
\end{align}
}

The functions $c_1(x)$ and $c_2(x)$ are defined as

\begin{align}
c_{1}(x)  &=\left(  1+x\right)  \ln\left(  1+x\right)
-x\ln\left(  x\right) \\
c_{2}(x) &=\left(  1+x\right)  \left[  \ln\left(
\frac{1+x}{x}\right)  \right]  ^{2}-\left(  \ln x\right)  ^{2}-2Li_{2}\left(-x\right). 
\end{align}

Here, $Li_{2}\left(  x\right)  $ is Spence's dilogarithm
function given by

\begin{align}
Li_{2}\left(  x\right)  &=\sum_{n=1}^{\infty}\frac{x^{n}}{n^{2}}\text{ \ \ ,
\ \ }\left\vert x\right\vert \leq1 \nonumber \\
Li_{2}\left(  -x\right)  &=\frac{\pi^{2}}{16}-\frac{1}{2}\left(  \ln x\right)
^{2}+\sum_{n=1}^{\infty}\frac{\left(  -1\right)  ^{n-1}}{n^{2}x^{n}}\text{
\ \ , \ \ }x>1
\end{align}

We can predict the behavior of the transport coefficients for high temperatures noting that, for $k_{B}T\gg J_{1}$, the Bose-Einstein function can be approximated by $n_{\lambda,k} \approx k_{B}T/E_{\lambda,k}$. We also note that, when $x\rightarrow \infty$, the functions $c_i(x)$ behave as \cite{Samajdar2019}:

\begin{align}
c_1(x)&=1+ln(x) \nonumber \\
c_2(x) &= \frac{\pi^2}{3}-\frac{1}{x}
\end{align}

{\color{black}
If $\Omega^\alpha_k= \pm \Omega^\beta_k$ and the Chern number of the bands is zero, we can show that

\begin{align}
\sigma_{xy}(T \rightarrow \infty) &= -\frac{k_B \, T}{\hbar^2}  \int\limits_{BZ} \frac{d^2k}{\left(  2\pi\right)  ^{2}} 
\left(\frac{1}{\omega^\alpha_k } \pm \frac{1}{\omega^\beta_k} \right) \Omega^\alpha_k \nonumber \\
\alpha_{xy}(T \rightarrow \infty) &= \frac{k_B}{\hbar}  \int\limits_{BZ} \frac{d^2k}{\left(  2\pi\right)  ^{2}} 
\,\, \left( ln \, \omega^\alpha_k \mp ln \, \omega^\beta_k \right) \Omega^\alpha_k \nonumber \\
\kappa_{xy}(T \rightarrow \infty) &= k_B  \int\limits_{BZ} \frac{d^2k}{\left(  2\pi\right)  ^{2}} 
\,\, \left( \omega^\alpha_k  \pm \omega^\beta_k   \right) \Omega^\alpha_k \label{highT}
\end{align}

Appendix B shows that the Berry curvatures have the same sign for the Union Jack lattice and opposite signs for the brick-wall lattice, so the equations above are valid for both systems studied here. The results above show that in the high-temperature limit, $\alpha_{xy}$ and $\kappa_{xy}$ are constants (asymptotic behavior), while $\sigma_{xy}$ is proportional to $T$.
}

\section{Brick-wall lattice}

\begin{figure}[h!]
\centering
\includegraphics[width=0.5\textwidth]{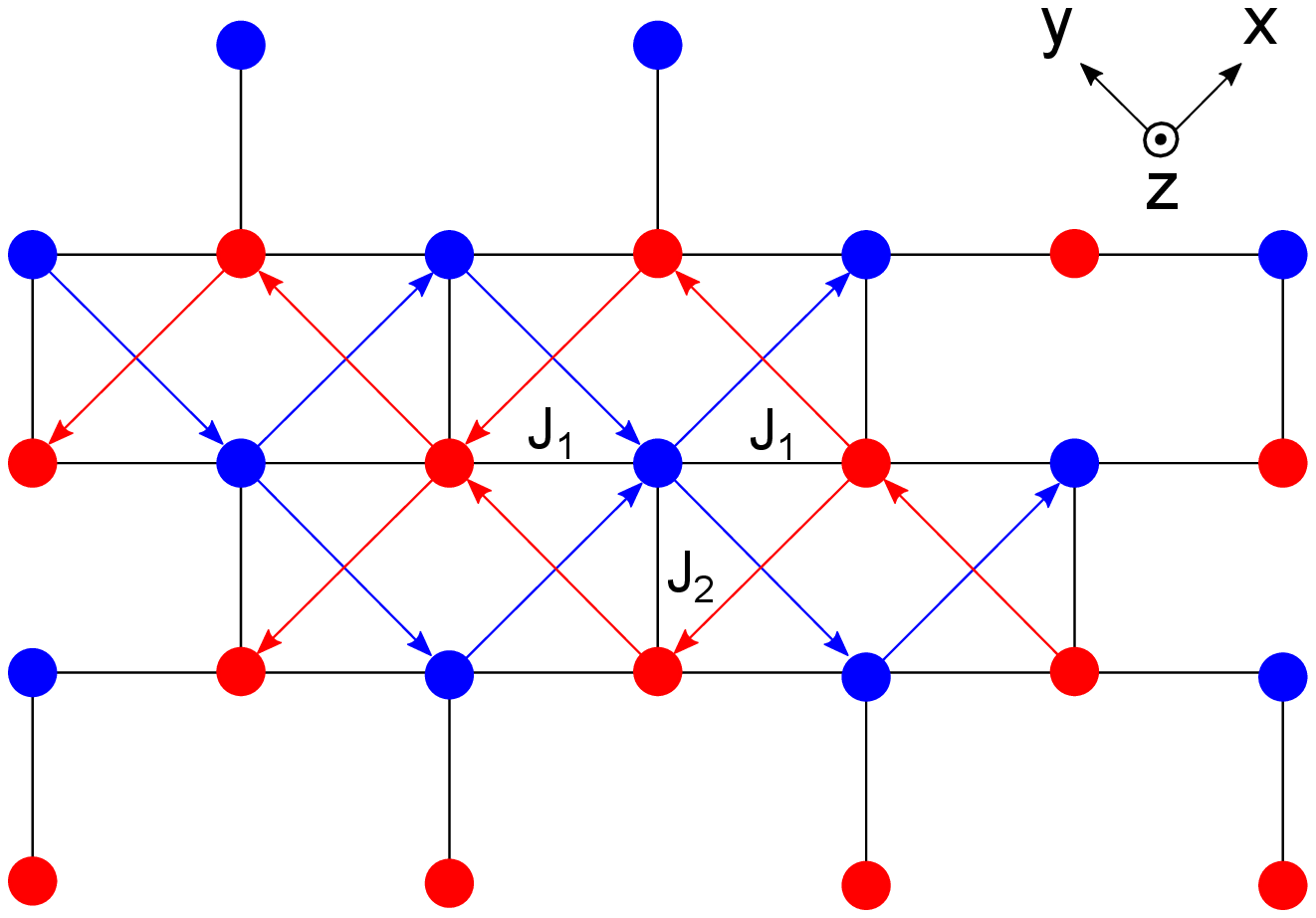} \caption{The AF brick-wall lattice described by Hamiltonian (\ref{brickwall}). The arrows represent DM interactions between NNN.} \label{fig_brickwall}%
\end{figure}

In the Union Jack lattice, all transport coefficients are non-null. Below, we will study a model that displays a magnonic equivalent of the QSHE, motivated by the fact that the prediction of this effect aroused a series of theoretical generalizations under various physical contexts. First, we analyze the brick-wall lattice, which can be considered a distorted honeycomb lattice \cite{Hou2015}. In this geometry, we consider the following Hamiltonian:  

\begin{equation}
H    =J_{1}\sum_{\left\langle i,j\right\rangle }\mathbf{S}_{i}\cdot
\mathbf{S}_{j}+ J_{2}\sum_{\left\langle i,j\right\rangle }\mathbf{S}_{i}\cdot \mathbf{S}_{j} + D \sum_{\left\langle \left\langle i,i' \right\rangle \right\rangle } \nu_{ii'} \mathbf{\hat{z}} \cdot  \mathbf{S}_{i} \times \mathbf{S}_{i'}  -A_1\sum_{i}\left(  S_{i}^{z}\right)  ^{2}-A_2\sum_{j}\left(  S_{j}^{z}\right)  ^{2}
\label{brickwall}%
\end{equation}

where $i$ denotes a site in sublattice $A$, and $j$ in the sublattice $B$. As shown in Figure \ref{fig_brickwall}, we have exchange interactions $J_1$ and $J_2$ between sites A and B in a ``wall of bricks" pattern. The arrows show the Dzyaloshinskii–Moriya interaction between sites AA and BB. We extend the model by considering different on-site anisotropies $A_1$ and $A_2$. In an experimental setup, that can be brought about by placing the sample on a substrate or heterostructure, thereby producing a local environment of the atoms that differ for the two sublattices. In this case, the non-magnetic atoms (which are responsible for $A_1 \neq A_2 \neq 0$) break the effective time-reversal symmetry \cite{Hidalgo-Sacoto2020}.

\begin{figure}[h!]
\centering
\includegraphics[width=0.5\textwidth]{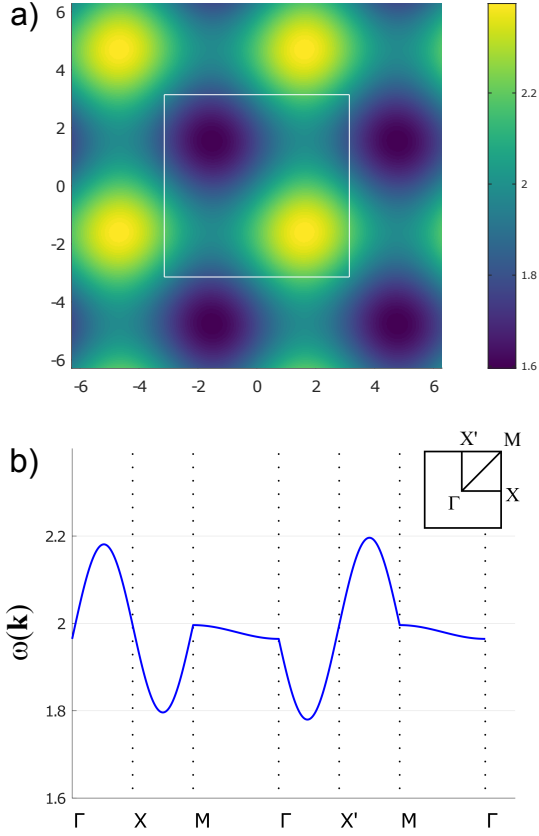} \caption{\textcolor{black}{Degenerate energy bands of the AF brick-wall lattice (a) in the Brillouin zone (white square) and (b) between the high symmetry points. Theory parameters are $S = J_1 = A_1 = A_2 = \alpha = 1$, and $D = 0.2$. The bands are degenerate for $A_1=A_2$.}} \label{bands_bwcomplete}%
\end{figure}

As before, we set the spacing of each sublattice equal to 1, and take the directions $\hat{\mathbf{x}}$ and $\hat{\mathbf{y}}$ along the diagonal axes of the two sublattices.
     
Following the same procedure as in the case of the Union Jack lattice, we get:

\begin{align}
H_{1} =&\frac{\left( 2+\alpha \right) }{2}J_{1}S\sum_{k}\left[a_{k}^{\dagger }a_{k}+a_{k}a_{k}^{\dagger }+b_{k}^{\dagger}b_{k}+b_{k}b_{k}^{\dagger }\right] +  \notag \\
&+\frac{3}{2}J_{1}S\sum_{k}\left[ \gamma _{k}^{\ast }\left(
a_{k}b_{-k}+b_{-k}a_{k}\right) +\gamma _{k}\left( a_{k}^{\dagger
}b_{-k}^{\dagger }+b_{-k}^{\dagger }a_{k}^{\dagger }\right) \right] \label{bw_xc}
\end{align}

\begin{align}
H_{SIA} =&\frac{1}{2}\left( 2S-1\right) \sum_{k} \left[ A_{1} \left( a_{k}^{\dagger}a_{k}+a_{k}a_{k}^{\dagger }\right)  
+A_{2} \left( b_{k}^{\dagger}b_{k}+b_{k}b_{k}^{\dagger }\right) \right]
\label{bw_sia}
\end{align}

\begin{align}
H_{DM}& =2SD\sum_{k}m_{k}\left[ \left( a_{k}^{\dagger
}a_{k}+a_{k}a_{k}^{\dagger }\right) -\left( b_{k}^{\dagger
}b_{k}+b_{k}b_{k}^{\dagger }\right) \right]
\label{bw_dmi}
\end{align}

where

\begin{align}
\alpha&=\frac{J_2}{J_1}, \ \ \ \ \ \gamma_k = \delta_k + i \varepsilon_k  \nonumber \\
\delta _{k} &=\frac{1}{12}\left[ \left( 2+\alpha \right) \cos \left( \frac{k_{x}}{2}\right) \cos \left( \frac{k_{y}}{2}\right) +\alpha \sin \left( 
\frac{k_{x}}{2}\right) \sin \left( \frac{k_{y}}{2}\right) \right],  \nonumber \\
\varepsilon _{k} &=-\frac{1}{12}\left[ \left( 2-\alpha \right) \cos \left( 
\frac{k_{x}}{2}\right) \sin \left( \frac{k_{y}}{2}\right) +\alpha \sin
\left( \frac{k_{x}}{2}\right) \cos \left( \frac{k_{y}}{2}\right) \right], \nonumber \\
m_{k}& =\frac{1}{2}\left( \sin k_{x}-\sin k_{y}\right).
\end{align}

Note that $m_k=-m_{-k}$. We have now the Hamiltonian in Eq. (\ref{hamilt_matrix}) with parameters:

\begin{align}
r_{1k} &=\left( \frac{2+\alpha }{2}\right) J_{1}S+2SDm_{k}+\frac{1}{2}%
A_{1}\left( 2S-1\right)  \notag \\
r_{2k} &=\left( \frac{2+\alpha }{2}\right) J_{1}S-2SDm_{k}+\frac{1}{2}%
A_{2}\left( 2S-1\right)  \notag \\
h_{xk} &=\frac{3}{2}J_{1}S\,\delta _{k}  \notag \\
h_{yk} &=-\frac{3}{2}J_{1}S\,\varepsilon _{k} \label{parameters_bw}
\end{align}%

{\color{black}
The imaginary part of a Bogoliubov-de Gennes Hamiltonian is represented by $h_{yk}$ (Eq. (\ref{M})). From the equations above, we see that $h_{yk} \propto \varepsilon _{k} = Im \, (\gamma_k) $. We also show in Appendix B that $h_{yk} \neq 0$ is a necessary condition for the Berry curvature to be non-zero. Hence, the complex lattice structure factor $\gamma_k$ is responsible for the non-zero Berry curvature in this system.

Another feature of this system is that the Berry curvature is independent of the DMI. Retrieving the definitions $f \equiv h_x+ih_y$ and $w \equiv \sqrt{r^2- |f|^2}$, from Eq. (\ref{BerryCurvature}) we can see that the Berry curvature ultimately depends on $h_x$, $h_y$ and $r$. As $r$ is defined as $r \equiv (r_1+r_2)/2$, from Eqs. (\ref{parameters_bw}) it is clear that neither $h_x$, $h_y$ or $r$ carry the DMI parameter ($D$ cancels out in $r$). Hence, the Berry curvature does not depend on $D$. This result was also found for the AF honeycomb lattice \cite{Cheng2016,Zyuzin2016,Owerre8,Zhang2022}.  
}

{\color{black}
The energy spectrum of the system is given by Eqs. (\ref{spectrum}) with

\begin{align}
w(\mathbf{k})  &  =\sqrt{r^{2}-\left\vert f\right\vert ^{2}} \nonumber \\
\Delta(\mathbf{k}) &=\frac{1}{4}\left(A_{1}-A_{2}\right) \left(2S-1\right)+ 2SDm_{k} \label{Delta_BW}
\end{align}

Let us first consider the case $A_1=A_2$. We have $\Delta(-\mathbf{k})=-\Delta(\mathbf{k})$, which means $\omega_\beta(\mathbf{k})=w(-\mathbf{k})-\Delta(-\mathbf{k})=w(\mathbf{k})+\Delta(\mathbf{k})=\omega_\alpha(\mathbf{k})$, and the bands are totally degenerate (Figure \ref{bands_bwcomplete}). This degeneracy is responsible for a pure spin Nernst effect of magnons, when a thermal gradient generates a transverse spin current without a net heat flow. This will be shown below.

Whereas the spin Hall and thermal Hall conductivities rely on ETRS
breaking, the spin Nernst effect does not \cite{Zhang2022}, and can exist even when the symmetry is present. Suppose the system shows ETRS. As we have seen, this implies $\Omega^\lambda\left(-\mathbf{k}\right) = -\Omega^\lambda \left(\mathbf{k} \right)$. In Appendix B we show that, $\Omega^\beta \left(\mathbf{k}\right) = \Omega^\alpha \left(\mathbf{-k}\right)$. These two properties imply $\Omega^\beta \left(\mathbf{k}\right)=-\Omega^\alpha \left(\mathbf{k}\right)$, i.e., the bands have Berry curvatures of opposite sign. Degenerate bands with opposite Berry curvatures was also predicted for the honeycomb AF lattice in Ref. \cite{Zyuzin2016}. However, we note that in Ref. \cite{Zhang2022} it is mentioned that there is another basis convention that leads to non-degenerate bands and same-sign Berry curvatures. This does not impact the results in transport coefficients. 

Considering the band degeneracy, the integrals in Eqs. (\ref{cond}) and (\ref{kappa}) can be developed as:

\begin{align}
\int d^2k \, \left[ f(\omega^\alpha_k) \Omega^\alpha_k  + f(\omega^\beta_k)\Omega^\beta_k \right] &= \int d^2k \, \left[ f(\omega^\alpha_k) - f(\omega^\beta_k) \right] \Omega^\alpha_k =0
\end{align}

The same does not occur for $\alpha_{xy}$, as there is a subtraction in the integrand of Eq. (\ref{alpha}), instead of a sum. Phenomenologically, that is the case presented in Section 2, where a thermal gradient $\partial_x T$ drives the two magnon modes in opposite transverse directions with the same intensity, making $\kappa_{xy}=0$ and $\alpha_{xy} \neq 0$. That is a magnonic equivalent of the quantum spin Hall effect of electrons. We call that phenomenon a pure spin Nernst effect of magnons, when a temperature gradient generates a spin current without heat flow. A field gradient $\partial_x B$, on the other hand, drives the two magnon modes in the same direction with the same intensity, so the spin current vanishes, $\sigma_{xy}=0$, while a heat current is present.

In the less strict case $A_1 \neq A_2$, the bands are not degenerate. Effective time-reversal symmetry is broken and all three transport coefficients can be non-null.

This brick-wall lattice is an example of the case discussed in Section 2 where the presence of an electric field (represented here by the DM term) is responsible for transverse transport of magnetic particles. This lattice has been proposed to describe the high-temperature superconductor $Ba_2CuO_{3+\delta}$ \cite{Wang2020}.}

\section{Results}

\subsection{Union Jack lattice}

\begin{figure}[h!]
\centering
\includegraphics[width=0.6\textwidth]{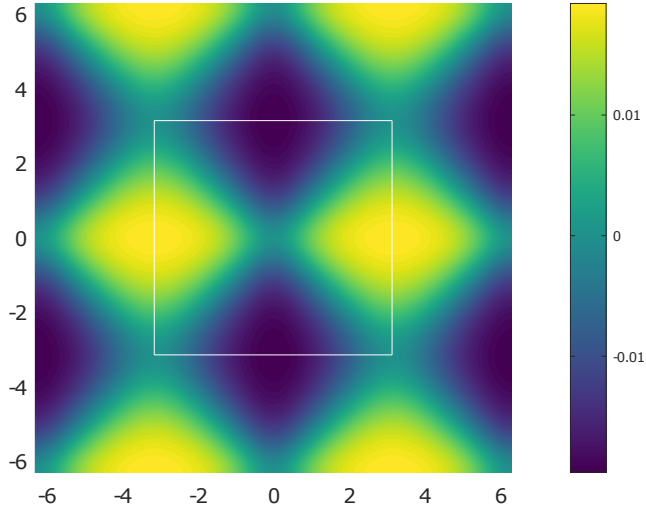}\caption{\textcolor{black}{Berry curvature of AF Union Jack lattice magnon bands. Both bands have the same Berry curvature (see discussion in Appendix B). The parameters are $S=J_{1}=A=1$, $J_{2}=D=0.2$, and $\lambda=\alpha=1.1$, which corresponds to the LSW bands in Figure \ref{bands_MSW} (solid red lines). The solid square is
the Brillouin zone.}} \label{berry_UJ}
\end{figure}

We now show the results of transversal transport for the antiferromagnetic Union Jack lattice, presented in Section 5. The results for energy dispersion for a variety of different parameters were already shown and discussed in the previous section. We set the parameters so that the Néel order is preserved, the Berry curvature is well-behaved, and the transport coefficients are non-null. The Berry curvature for typical parameters is represented in Figure \ref{berry_UJ}. We see that the Berry curvature has the property $\Omega\left(\mathbf{k}\right) = \Omega \left( -\mathbf{k} \right)$, which comes from the space inversion symmetry of the magnetic lattice (this symmetry also gives rise to the property $\omega_{\alpha,\beta}\left(\mathbf{k}\right) = \omega_{\alpha,\beta} \left( -\mathbf{k} \right)$). \textcolor{black}{For this system, ETRS is broken, but the Chern number is still zero:}

\begin{figure}[h!]
\centering
\includegraphics[width=0.4\textwidth]{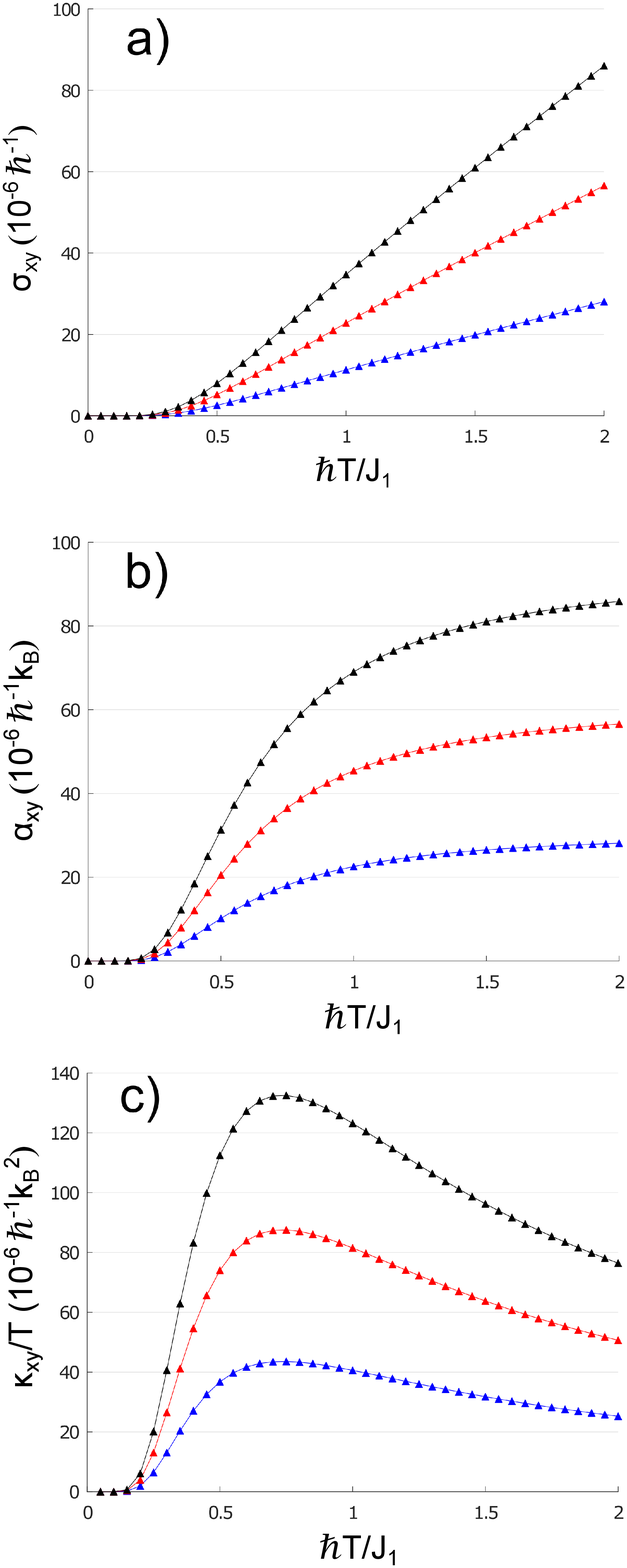}\caption{Transport coefficients of the AF Union Jack lattice for $D=0.1$ (blue), $D=0.2$ (red) and $D=0.3$ (black). The other theory parameters are $S=J_{1}=A=1$, $J_{2}=0.2$, $\lambda=\alpha
=1.1$. (a) Spin Hall conductivity. (b) Spin Nernst coefficient. (c) Thermal Hall
conductivity \textcolor{black}{over T}.} \label{transport_UJ_complete}
\end{figure}

\begin{figure}[h!]
\centering
\includegraphics[width=0.4\textwidth]{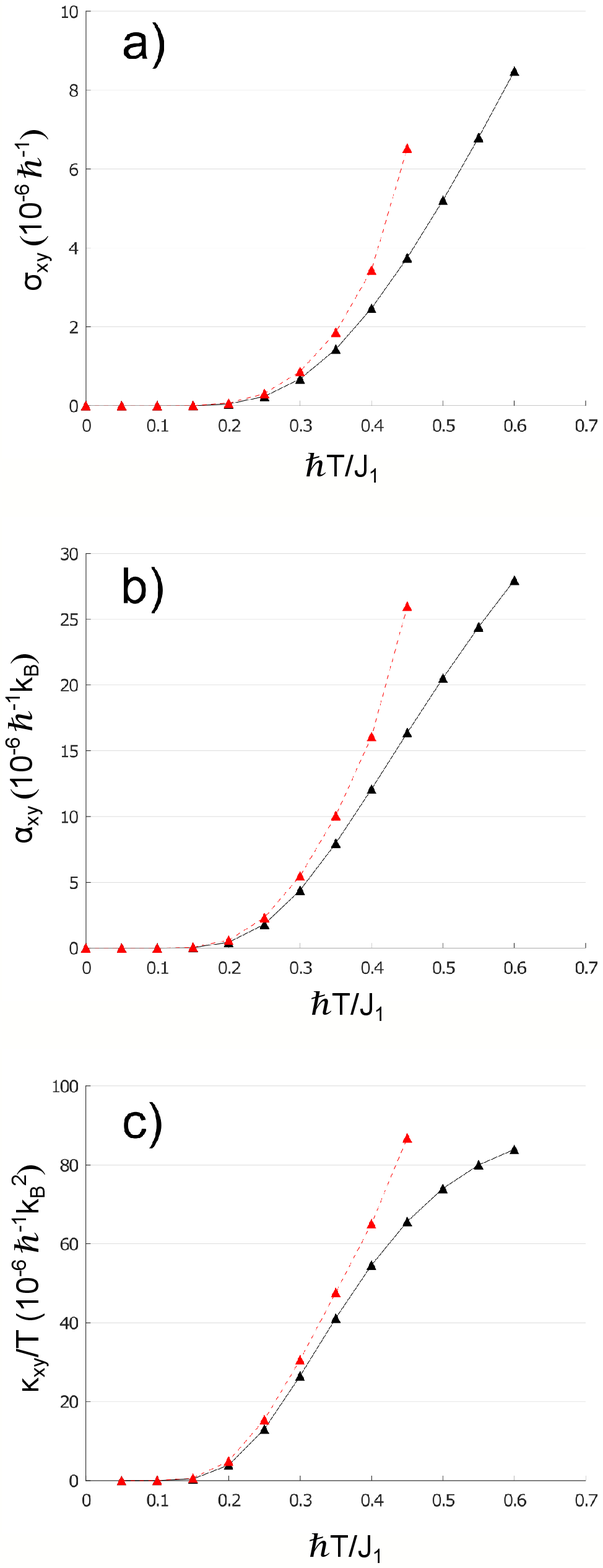} \caption{Effects of the MSW in the Union Jack lattice transport coefficients. (a) Spin Hall
conductivity, (b) spin Nernst
coefficient and (c) thermal Hall
conductivity \textcolor{black}{over T} for the LSW (black) and MSW (dashed red) when
anharmonic terms are included with a mean-field approach. The theory parameters are
$S=J_{1}=A=1$, $J_{2}=D=0.2$, $\lambda=\alpha=1.1$.} \label{coef_MSW}
\end{figure}

\begin{equation}
C_{\alpha,\beta}= \int \frac{d^2k}{2 \pi} \Omega (\mathbf{k}) =0
\end{equation}

That happens for every combination of Hamiltonian parameters, meaning the system is always topologically trivial and does not present protected edge states. That, however, does not necessarily means the transport coefficients are null, as the integrands in Eqs. (\ref{cond})-(\ref{kappa}) are weighted by functions of the energies $f(\omega_{\alpha,\beta}(\mathbf{k}))$. The in-plane anisotropy $\alpha \neq 1$ generates an energetic imbalance between bands so that the integrals are found to be non-null. Although they are non-null, it is known that the transport coefficients for systems with $C=0$ are much smaller compared to the cases where $C \neq 0$ \cite{Samajdar2019}.

To have a non-null Berry curvature, we need an imaginary term in the Hamiltonian ($h_y \neq 0$). As $h_y$ comes from the DM interaction, we need $D \neq 0$: the DMI is crucial for a non-zero Berry curvature for the Union Jack lattice. That contrasts with systems where the DM
interaction happens between sites of the same type (i. e., AA and BB) and
contributes with a real term in the Hamiltonian, as it is the case of the honeycomb and brick-wall lattices. Also, we need $A \neq 0$ and $S \neq 1/2$ for non-zero energy in both bands gap, preventing the magnon occupation number (and the Berry curvature) from diverging when $T>0$.

First, we present results for transport coefficients using the linear spin wave formalism (LSW). \textcolor{black}{In all plots, we set $\hbar= k_B=1$, so the units of the transport coefficients are the constants in front of the integrals in Eqs. (\ref{cond}-\ref{kappa}).} In Figure \ref{transport_UJ_complete}(a) we show the spin Hall conductivity $\sigma_{xy}$ as a function of $T$ for $S=J_1=A=1$, $J_2=0.2$, $\lambda=\alpha=1.1$ and three values of $D$. At $T = 0$, $\sigma_{xy}$ (and all other transport coefficients) vanishes due to the absence of magnon excitations. Magnons are thermally excited as the temperature increases, and the transport coefficients become finite. In Figures \ref{transport_UJ_complete}(b) and \ref{transport_UJ_complete}(c) we present the Nernst coefficient and thermal Hall conductivity versus temperature for the same values of parameters. \textcolor{black}{Focusing in the thermal Hall conductivity $\kappa_{xy}$ versus $T$, the behavior of the curve resembles several other magnetic systems, like the FM and non-collinear AFM honeycomb lattice \cite{Owerre3,Owerre8}, AFM checkerboard lattice \cite{Pires2020}, and FM kagome lattice, both in theoretical \cite{Laurell2018} and experimental \cite{Doki2018} studies. For some systems, the sign of $\kappa_{xy}$ can change with the choice of parameters (like in the Kitaev model \cite{Li2022}). In others, the curve $\kappa_{xy}$ versus $T$ can show peaks or valleys and even change sign with the temperature increase \cite{Mook2014,Cao2015}. This behavior was not observed in the AFM Union Jack lattice, where the curve $\kappa_{xy}$ versus $T$ is monotonic and asymptotic.}

As we can see, all transport coefficients increase with D. If $J_2=0$ or $\alpha=1$, a symmetry of the energy bands $\omega_{\alpha,\beta}(\mathbf{k})$ (namely, a rotation of the energy functions by $90^\circ$) makes all coefficients vanish identically. If $J_2\neq 0$ and $\alpha \neq 1$, this symmetry is broken by an energetic inequivalence (mainly in the lower band, see Figure \ref{bands_alpha}) of the axes $k_x$ and $k_y$. That generates transverse transport. The effect of the off-plane exchange anisotropy $\lambda$ is only quantitative: transverse transport exists even when $\lambda=1$. Also, we can see that the high-temperature behavior of the coefficients is in accordance to Eqs. (\ref{highT}): $\alpha_{xy}$ and $\kappa_{xy}$ have asymptotic behavior, while $\sigma_{xy}$ is proportional to $T$.

Now, we consider the effect of anharmonic contributions using a modified spin wave theory, as mentioned in Section 5. In the linear spin wave treatment, the temperature dependence of the transport coefficients comes only from the Bose-Einstein distribution $n_k^{\alpha,\beta}$, the energy bands and Berry curvature being independent of temperature. In the self-consistent MSW theory, we partly include the effects of finite $T$ and quantum fluctuations with a mean-field approach. We present the calculations in Appendix A. The corrections make the energy dispersion and Berry curvature temperature-dependent, correcting the transport coefficients for each temperature. In Figure \ref{coef_MSW} we present the results for MSW. The parameters are $S=J_1=A=1$, $J_2=D=0.2$, $\lambda=\alpha=1.1$. The self-consistent corrections lower the energy bands (see Figure \ref{bands_MSW_T}), raising the magnon population for a given temperature and increasing all transport coefficients. That happens until we reach a temperature $T_N$ ($ \approx 0.456$ for the chosen parameters), where no self-consistent solution is found anymore. At this point, the magnetization vanishes (see Figure \ref{magn}) signaling a transition to a disordered phase.

\begin{figure}[h!]
\centering
\includegraphics[width=0.6\textwidth]{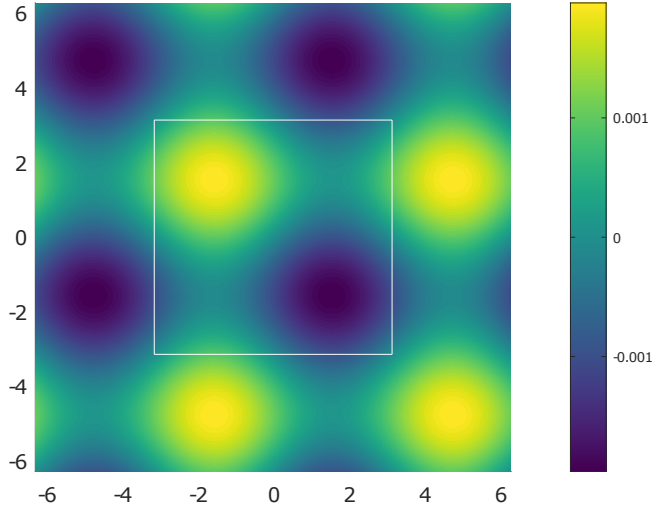} \caption{\textcolor{black}{Berry curvature of band $\alpha$ for the AF brick-wall lattice (the band structure is shown in Figure \ref{bands_bwcomplete}). The parameters are $S=J_1=A_1=A_2=\alpha=1$ and $D=0.2$. The band $\beta$ has Berry curvature of opposite sign ($\Omega_\beta (\mathbf{k})=-\Omega_\alpha (\mathbf{k})$), as discussed in Appendix B.}} \label{berry_bw}%
\end{figure}

\begin{figure}[h!]
\centering
\includegraphics[width=0.5\textwidth]{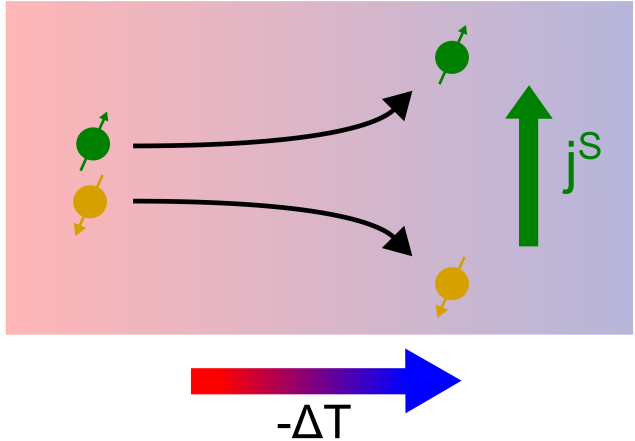} \caption{The pure spin Nernst effect of magnons, when different magnons are deflected on opposite directions with the same intensity, so it shows a spin current with no net heat flow.} \label{SNE}%
\end{figure}

\begin{figure}[h!]
\centering
\includegraphics[width=0.5\textwidth]{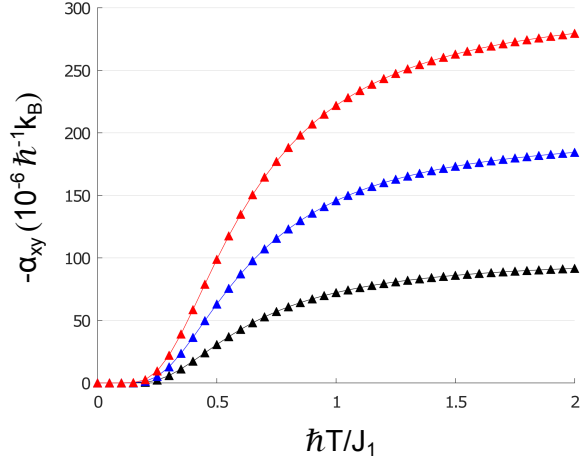} \caption{Spin Nernst coefficient as a function of temperature for the AF brick-wall lattice. The parameters are $S=J_1=A_1=A_2=\alpha=1$ and three different values of DM parameter: $D=0.1$ (black), $D=0.2$ (blue) and $D=0.3$ (red). For $A_1=A_2$, we have $\sigma_{xy}=\kappa_{xy}=0$, and the system shows a pure spin Nernst effect of magnons.} \label{nernst_bw}%
\end{figure}

\begin{figure}[h!]
\centering
\includegraphics[width=0.7\textwidth]{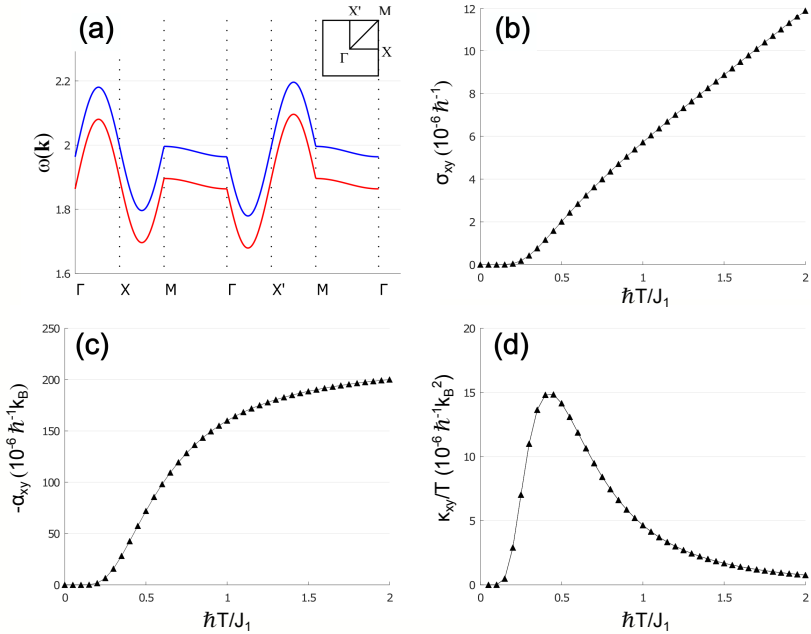} \caption{(a) Energy bands and (b-d) transport coefficients of the AF brick-wall lattice when $A_1 \neq A_2$. \textcolor{black}{The parameters are $S=J_1=\alpha=A_1=1$, $A_2=0.8$, and $D=0.2$}. All three coefficients are non-null.} \label{bw_SIAdif}%
\end{figure}

\subsection{Brick-wall lattice}

\textcolor{black}{We turn to the antiferromagnetic brick-wall lattice, firstly analyzing the case where $A_1=A_2$, when the bands are degenerate. Without DMI ($D=0$) the bands present even parity: $\omega_\beta(\mathbf{k})=\omega_\alpha(\mathbf{k})=\omega_\alpha(\mathbf{-k})$.} The Hamiltonian $H=H_J+ H_{SIA}$ is invariant under the effective time-reversal symmetry $\mathcal{T}c_x$, as shown in Section 6. The presence of an ETRS results in an odd Berry curvature $\Omega \left( \mathbf{k} \right) = -\Omega \left( -\mathbf{k} \right)$. Just like the case of the honeycomb lattice \cite{Zyuzin2016,Zhang2022}, a nonzero Berry curvature develops even without the DM interaction, but all transport coefficients are null. This non-zero Berry curvature also means that $H_J$ must break the inversion symmetry. \textcolor{black}{If we now introduce the DM interaction, the ETRS is broken in the Hamiltonian, but that does not affect the eigenstates}. The Berry curvature is unaffected and holds the odd parity (Figure \ref{berry_bw}). In a certain way, we can say that the ETRS is still present in the eigenstates of the system. As in the QSHE for electrons (mentioned in the Introduction), ETRS implies $C = 0$, and we have edge modes protected by the effective time-reversal symmetry. ETRS can protect gapped band topology and is associated with a two-valued or $\mathbb{Z}_2$ topological index \cite{Kondo2019}. \textcolor{black}{For $D \neq 0$, we have $\sigma_{xy}=\kappa_{xy}=0$ and $\alpha_{xy} \neq 0$, as discussed in Section 8}. That happens even in the case of anisotropic exchange, when $\alpha=J_2/J_1 \neq 1$. The anisotropy $\alpha$ only changes quantitatively the spin Nernst coefficient.

\textcolor{black}{In other words: when $A_1=A_2$ and $D = 0$, the bands are totally degenerate and even, but $\Omega \left( \mathbf{k} \right)$ is odd. Thus, all transport coefficients vanish. A non-null $D$ breaks the even parity of the bands.} In the presence of a temperature gradient $\partial_xT$, the effect of the DM term is opposite for the two modes, driving the up/down magnons in opposite transverse directions with the same intensity. The transverse thermal current vanishes identically, but we have a net spin current ($\kappa_{xy}=0$ and $\alpha_{xy} \neq 0$). This is the pure spin Nernst effect of magnons (Figure \ref{SNE}). When subjected to a magnetic field gradient $\partial_xB$, both magnon modes are driven to the same transverse direction with the same intensity, generating a heat current with no net spin current ($\sigma_{xy}=0$).

The results for the spin Nernst coefficient as a function of temperature are shown in Figure \ref{nernst_bw}. The parameters are $S=J_1=A_1=A_2=\alpha=1$, and three different values of $D$. We can see that $\alpha_{xy}$ increases with $D$, and has an asymptotic behavior in the high-temperature limit, as predicted.

\textcolor{black}{All three transport coefficients are non-null for the general case $A_1 \neq A_2$. When subjected to a thermal gradient, up/down magnons are driven in opposite directions, but now with different intensities. The system shows non-null thermal and spin currents. The same can be said about the response to a magnetic field gradient, except now the magnons flow in the same direction. We show the results for this case in Figure \ref{bw_SIAdif}.}

\section{Conclusions}

In summary, we discussed magnon transport in antiferromagnetic topological insulators. We treated the antiferromagnetic Union Jack and the brick-wall lattices as examples, but the methods discussed here could be applied to several other lattices, such as the staggered and zig-zag square lattices \cite{Kondo2022}.

{\color{black}
Regarding the AF Union Jack lattice, we studied the band structure behavior for several parameters. The DMI generates a Berry curvature. It is an even function $\Omega(\mathbf{k})=\Omega(\mathbf{-k})$ and identical for the two magnons. The bands are not degenerate, resulting in non-zero transport coefficients. Including magnon-magnon interactions with a mean-field approach lowers the energy bands, raising the thermal population. As a consequence, transverse transport is intensified. 

Turning to the AF brick-wall lattice, the system presents degenerate bands with opposite Berry curvature. The Berry curvature is an odd function ($\Omega(\mathbf{k})=-\Omega(\mathbf{-k})$), is independent of the DMI and relies on a complex structure factor. The system can present a pure spin Nernst effect, where $\alpha_{xy} \neq 0$ and $\sigma_{xy}=\kappa_{xy}=0$. This effect is protected by an effective time-reversal symmetry, which can be broken by different single-ion anisotropies for the two sublattices. In this case the bands split, and all three transport coefficients are non-null.}

We have shown that there is a fundamental difference in the microscopic origin of the spin Nernst effect on the two lattices. Although the Dzyaloshinskii-Moriya interaction (DMI) is an essential tool in both cases, it plays an entirely different role at the microscopic level. For example, the DMI induces a nonzero Berry curvature in the Union Jack lattice but not in the brick-wall lattice.  

As far as we know, up to now, there are no real systems described by the Union Jack lattice. However, Wioland et al \cite{Wioland2016} found that lattices of hydrodynamically coupled bacterial vortices can spontaneously organize into distinct patterns characterized by ferro and antiferromagnetic order in a Union Jack lattice. They also found the existence of geometry-induced edge currents reminiscent of those in the quantum Hall effect. The  Union Jack lattice could also be realized in optical lattices where synthetic DM interaction can be generated using laser beams \cite{Olschlager2012}. In the next step, we intend to study the dynamics of the model following Refs. \cite{Azizi2020,AziziPrePrint}.

\section{Acknowledgments}

This work was supported by CAPES (Coordenação de Aperfeiçoamento de Pessoal de Nível Superior) and CNPq (Conselho Nacional de Desenvolvimento Científico e Tecnológico).



\begin{appendices}
\section{Modified Spin Wave (MSW) approach}
\renewcommand{\theequation}{\thesection{.}\arabic{equation}}
\setcounter{equation}{0}

To calculate the effect of the lowest-order anharmonic contributions to the Union Jack lattice, presented in Section 5, we start
with the Holstein-Primakoff transformation written up to four operator terms:

\begin{equation}
S_{i}^{+}=\sqrt{2S}\left(  a_{i}-\frac{a_{i}^{\dagger}a_{i}a_{i}}{4S}\right)
,\text{ \ \ }S_{i}^{-}=\sqrt{2S}\left(  a_{i}^{\dagger}-\frac{a_{i}^{\dagger
}a_{i}^{\dagger}a_{i}}{4S}\right)  ,\text{ \ \ }S_{i}^{z}=\,S-a_{i}^{\dagger
}a_{i} \label{A1}%
\end{equation}

on a sublattice A and

\begin{equation}
S_{j}^{+}=\sqrt{2S}\left(  b_{j}^{\dagger}-\frac{b_{j}^{\dagger}b_{j}%
^{\dagger}b_{j}}{4S}\right)  \,,\ \ S_{j}^{-}=\sqrt{2S}\left(  b_{j}%
-\frac{b_{j}^{\dagger}b_{j}b_{j}}{4S}\right)  ,\ \ S_{j}^{z}=\,-S+b_{j}%
^{\dagger}b_{j} \label{A2}%
\end{equation}

on sublattice \textit{B}. Taking (\ref{A1}) and (\ref{A2}) into Hamiltonian
(\ref{hamilt}) we find, neglecting all constant terms:

\begin{equation}
H=H_{1}+H_{2}+H_{DM}+H_{SIA}
\end{equation}

where

\begin{align}
H_{1}    &= J_{1}S\sum_{\left\langle i,j\right\rangle }\left(  a_{i}^{\dagger
}a_{i}+b_{j}^{\dagger}b_{j}+a_{i}b_{j}+a_{i}^{\dagger}b_{j}^{\dagger}\right)
+\nonumber\\
&  -\frac{J_{1}}{4}\sum_{\left\langle i,j\right\rangle }\left(  a_{i}%
b_{j}^{\dagger}b_{j}b_{j}+a_{i}^{\dagger}a_{i}a_{i}b_{j}+a_{i}^{\dagger}%
b_{j}^{\dagger}b_{j}^{\dagger}b_{j}+a_{i}^{\dagger}a_{i}^{\dagger}a_{i}%
b_{j}^{\dagger}+4a_{i}^{\dagger}a_{i}b_{j}^{\dagger}b_{j}\right) \\
H_{2}    &= S\sum_{\left\langle \left\langle i,j\right\rangle \right\rangle
}J_{2,ij}\left[  a_{i}a_{j}^{\dagger}+a_{i}^{\dagger}a_{j}-\lambda\left(
a_{i}^{\dagger}a_{i}+a_{j}^{\dagger}a_{j}\right)  \right]  +\nonumber\\
&  -\frac{1}{4}\sum_{\left\langle \left\langle i,j\right\rangle \right\rangle
}J_{2,ij}\left(  a_{i}a_{j}^{\dagger}a_{j}^{\dagger}a_{j}+a_{i}^{\dagger}%
a_{i}a_{i}a_{j}^{\dagger}+a_{i}^{\dagger}a_{j}^{\dagger}a_{j}a_{j}%
+a_{i}^{\dagger}a_{i}^{\dagger}a_{i}a_{j}-4\lambda a_{i}^{\dagger}a_{i}%
a_{j}^{\dagger}a_{j}\right) \\
H_{DM}  &=   iDS \sum_{\left\langle i,j\right\rangle }\nu_{ij}\left(  a_{i}%
b_{j}-a_{i}^{\dagger}b_{j}^{\dagger}\right)  +\nonumber\\
&  +i\frac{D}{4}\sum_{\left\langle i,j\right\rangle }\nu_{ij}\left(
-a_{i}b_{j}^{\dagger}b_{j}b_{j}-a_{i}^{\dagger}a_{i}a_{i}b_{j}+a_{i}^{\dagger
}b_{j}^{\dagger}b_{j}^{\dagger}b_{j}+a_{i}^{\dagger}a_{i}^{\dagger}a_{i}%
b_{j}^{\dagger}\right)
\end{align}

\begin{align}
H_{SIA}  &  =2AS\left(  \sum_{i\in A}a_{i}^{\dagger}a_{i}+\sum_{j\in B}%
b_{j}^{\dagger}b_{j}\right)  -A\left(  \sum_{i\in A}a_{i}^{\dagger}a_{i}%
a_{i}^{\dagger}a_{i}+\sum_{j\in B}b_{j}^{\dagger}b_{j}b_{j}^{\dagger}%
b_{j}\right) \nonumber\\
&  =A\left(  2S-1\right)  \left(  \sum_{i\in A}a_{i}^{\dagger}a_{i}+\sum_{j\in
B}b_{j}^{\dagger}b_{j}\right)  -A\left(  \sum_{i\in A}a_{i}^{\dagger}%
a_{i}^{\dagger}a_{i}a_{i}+\sum_{j\in B}b_{j}^{\dagger}b_{j}^{\dagger}%
b_{j}b_{j}\right)  \label{SIA_2S1}%
\end{align}

where, in the $SIA$ term, we have used $a_{i}^{\dagger}a_{i}a_{i}^{\dagger
}a_{i}=a_{i}^{\dagger}a_{i}+a_{i}^{\dagger}a_{i}^{\dagger}a_{i}a_{i}$ (and
similarly for the $b_{j}$ operatores) to normal order the quartic terms.

Considering only the quadratic terms on the expressions above, we obtain the
linear spin wave theory (LSW) exposed in the main text. We include the quartic terms to consider
the interactions between magnons and perform a
mean-field decoupling to obtain an effective quadratic Hamiltonian (modified
spin theory, MSW). We use the well-known relation between quantum operators (ignoring the zeroth-order
terms $\left\langle AB\right\rangle \left\langle CD\right\rangle $ which only
add a global constant energy to the spectrum):

\begin{equation}
ABCD=\left\langle AB\right\rangle CD+AB\left\langle CD\right\rangle
+\left\langle AC\right\rangle BD+AC\left\langle BD\right\rangle +\left\langle
AD\right\rangle BC+AD\left\langle BC\right\rangle
\end{equation}

The only non-null mean-field terms are $\left\langle a_{i}^{\dagger}%
a_{j}\right\rangle $, $\left\langle b_{i}^{\dagger}b_{j}\right\rangle $,
$\left\langle a_{i}b_{j}\right\rangle $, $\left\langle b_{i}a_{j}\right\rangle
$ and their complex conjugates (see discussion in the end of this Appendix).
For instante, the first quartic term in $H_{1}$ decouples as:

\begin{equation}
a_{i}b_{j}^{\dagger}b_{j}b_{j}=2\left\langle b_{j}^{\dagger}b_{j}\right\rangle
a_{i}b_{j}+2\left\langle a_{i}b_{j}\right\rangle b_{j}^{\dagger}b_{j}%
\end{equation}

We rename the mean-field terms as

\begin{align}
g_{1}  &  =\left\langle a_{i}^{\dagger}a_{i}\right\rangle ,\text{ \ \ }%
g_{2}=\left\langle b_{j}^{\dagger}b_{j}\right\rangle ,\text{ \ \ }%
g_{3}=\left\langle a_{i}b_{j}\right\rangle ,\nonumber\\
g_{4}  &  =\left\langle a_{i}^{\dagger}b_{j}^{\dagger}\right\rangle ,\text{
\ \ }g_{5}=\left\langle a_{i}^{\dagger}a_{j}\right\rangle ,\text{ \ \ }%
g_{6}=\left\langle a_{i}a_{j}^{\dagger}\right\rangle
\end{align}

Noting that $g_{3}$ and $g_{4}$ can be complex, we can also write:%

\begin{align}
&  g_{3}=G_{1}+iG_{2},\text{ \ \ }g_{4}=G_{1}-iG_{2},\text{ \ \ }\nonumber\\
&  \rightarrow g_{3}+g_{4}=2G_{1},\text{ \ \ }g_{3}-g_{4}=2iG_{2}%
\end{align}

With these definitions, we write

\begin{align}
H_{1}  &  =J_{1}S\sum_{\left\langle i,j\right\rangle }\left(  a_{i}^{\dagger
}a_{i}+b_{j}^{\dagger}b_{j}+a_{i}b_{j}+a_{i}^{\dagger}b_{j}^{\dagger}\right)
-J_{1}\sum_{\left\langle i.j\right\rangle }\left[  \left(  g_{2}%
+G_{1}\right)  a_{i}^{\dagger}a_{i}+\left(  g_{1}+G_{1}\right)  b_{j}%
^{\dagger}b_{j}\right.  +  \\
&  \,\left.  +\left(  \frac{g_{1}+g_{2}}{2}+G_{1}\right)  \left(  a_{i}%
b_{j}+a_{i}^{\dagger}b_{j}^{\dagger}\right)  -iG_{2}\left(  a_{i}b_{j}%
-a_{i}^{\dagger}b_{j}^{\dagger}\right)  \right] \nonumber
\end{align}

\begin{align}
H_{2}  &  =S\sum_{\left\langle \left\langle i,j\right\rangle \right\rangle
}J_{2,ij}\left[  a_{i}a_{j}^{\dagger}+a_{i}^{\dagger}a_{j}-\lambda\left(
a_{i}^{\dagger}a_{i}+a_{j}^{\dagger}a_{j}\right)  \right]   -\sum_{\left\langle \left\langle i,j\right\rangle \right\rangle }%
J_{2,ij}\left[  \left(  \frac{g_{5}+g_{6}}{2}-\lambda g_{1}\right)  \left(
a_{i}^{\dagger}a_{i}+a_{j}^{\dagger}a_{j}\right)  \right.  +  \\
&  \left.  +g_{1}\left(  a_{i}^{\dagger}a_{j}+a_{i}a_{j}^{\dagger}\right)
-\lambda\left(  g_{6}a_{i}^{\dagger}a_{j}+g_{5}a_{i}a_{j}^{\dagger}\right)
\right] \\ \nonumber
H_{DM}  &  =iDS\sum_{\left\langle i,j\right\rangle }\nu_{ij}\left(  a_{i}%
b_{j}-a_{i}^{\dagger}b_{j}^{\dagger}\right)  +D \sum_{\left\langle i,j\right\rangle }\nu_{ij}\left[  G_2 \left(  a_{i}^{\dagger}a_{i}+b_{j}^{\dagger}b_{j}\right)
- \frac{i}{2}\left(  g_{1}+g_{2}\right)  \left(  a_{i}b_{j}-a_{i}^{\dagger}b_{j}^{\dagger
}\right)  \right]\\
H_{SIA}&=A\left(  2S-1\right)  \left(  \sum_{i\in A}a_{i}^{\dagger}a_{i}%
+\sum_{j\in B}b_{j}^{\dagger}b_{j}\right)  -4A\left(  \sum_{i\in A}g_{1}%
a_{i}^{\dagger}a_{i}+\sum_{j\in B}g_{2}b_{j}^{\dagger}b_{j}\right)
\end{align}

Evaluating:

\begin{align}
H_{1}+H_{DM}  &  =\sum_ {\left\langle i,j\right\rangle } \left\{ \left[  J_{1}\left(
S-g_{2}-G_{1}\right)  +\nu_{ij}DG_{2}\right]  a_{i}^{\dagger}a_{i}+\left[
J_{1}\left(  S-g_{1}-G_{1}\right)  +\nu_{ij}DG_{2}\right]  b_{j}^{\dagger
}b_{j}+  \right. \\
&  \,+J_{1}\left(  S-\frac{g_{1}+g_{2}}{2}-G_{1}\right)  \left(  a_{i}%
b_{j}+a_{i}^{\dagger}b_{j}^{\dagger}\right)  \left. +i\left[  J_{1}G_{2}+\nu_{ij}D\left(  S-\frac{g_{1}+g_{2}}{2}\right)
\right]  \left(  a_{i}b_{j}-a_{i}^{\dagger}b_{j}^{\dagger}\right) \right\}  \nonumber \\
H_{2}  &  =\sum_{\left\langle \left\langle i,j\right\rangle \right\rangle
}J_{2,ij}\left[  \left(  -\frac{g_{5}+g_{6}}{2}-\lambda\left(  S-g_{1}\right)
\right)  \left(  a_{i}^{\dagger}a_{i}+a_{j}^{\dagger}a_{j}\right)  \right. + \\
&\left.  +\left(  S-g_{1}\right)  \left(  a_{i}^{\dagger}%
a_{j}+a_{i}a_{j}^{\dagger}\right)  +\lambda\left(  g_{6}a_{i}^{\dagger}%
a_{j}+g_{5}a_{i}a_{j}^{\dagger}\right)  \right] \nonumber \\
H_{SIA} &=A\left[  \left(  2S-1\right)  -4g_{1}\right]  \sum_{i\in A}%
a_{i}^{\dagger}a_{i}+A\left[  \left(  2S-1\right)  -4g_{2}\right]  \sum_{j\in
B}b_{j}^{\dagger}b_{j}%
\end{align}

Let $\Gamma_{i}$ $\ $be new parameters defined as:

\begin{align}
\Gamma_{1}  &  =S-\left(  \frac{g_{1}+g_{2}}{2}+G_{1}\right) \nonumber\\
\Gamma_{2}  &  =\left(  S-G_{1}-g_{2}\right) \nonumber\\
\Gamma_{3}  &  =\left(  S-G_{1}-g_{1}\right) \nonumber\\
\Gamma_{4}  &  =S-\frac{g_{1}+g_{2}}{2}\nonumber\\
\Gamma_{5}  &  =-\frac{g_{5}+g_{6}}{2}-\lambda\left(  S-g_{1}\right)
\nonumber\\
\Gamma_{6}  &  =\left(  2S-1\right)  -4g_{1}\nonumber\\
\Gamma_{7}  &  =\left(  2S-1\right)  -4g_{2} \label{Gamma}%
\end{align}

We can rewrite the Hamiltonian in terms of eight temperature-dependent
parameters: $\Gamma_{i}$ and $G_{2}$ (not all linearly independent). Fourier
transforming and symmetrizing the operators, we get:

\begin{align}
H_{1}+H_{DM}  &  =2\sum_{k}\left[  \left(  J_{1}\Gamma_{2}+m_{k}DG_{2}\right)
\left(  a_{k}^{\dagger}a_{k}+a_{k}a_{k}^{\dagger}\right)  +\left(  J_{1}%
\Gamma_{3}+m_{k}DG_{2}\right)  \left(  b_{k}^{\dagger}b_{k}+b_{k}%
b_{k}^{\dagger}\right)  \right.  +\nonumber\\
&  \,+\gamma_{k}J_{1}\Gamma_{1}\left(  a_{k}b_{-k}+b_{-k}a_{k}+a_{k}^{\dagger
}b_{-k}^{\dagger}+b_{-k}^{\dagger}a_{k}^{\dagger}\right)  +\nonumber\\
&  +\left.  i\left(  \gamma_{k}J_{1}G_{2}+m_{k}D\Gamma_{4}\right)  \left(
a_{k}b_{-k}+b_{-k}a_{k}-a_{k}^{\dagger}b_{-k}^{\dagger}-b_{-k}^{\dagger}%
a_{k}^{\dagger}\right)  \right]
\end{align}
\bigskip%
\begin{equation}
H_{2}=J_{2}\sum_{k}\left\{  \Gamma_{5}\left[  \left(  \alpha+1\right)
-2\lambda\eta_{k}\right]  +2\eta_{k}\left(  1-\lambda^{2}\right)  \left(
\Gamma_{3}+\Gamma_{4}-\Gamma_{1}\right)  \right\}  \left(  a_{k}^{\dagger
}a_{k}+a_{k}a_{k}^{\dagger}\right)
\end{equation}

\begin{equation}
H_{SIA}=\frac{A}{2}\sum_{k}\left[  \Gamma_{6}\left(  a_{k}^{\dagger}%
a_{k}+a_{k}a_{k}^{\dagger}\right)  +\Gamma_{7}\left(  b_{k}^{\dagger}%
b_{k}+b_{k}b_{k}^{\dagger}\right)  \right]
\end{equation}

We have, then, a renormalized Hamiltonian matrix

\begin{equation}
H_{k}=\left(
\begin{array}
[c]{cc}%
M_{k} & 0\\
0 & M_{-k}^{\ast}%
\end{array}
\right)  \text{, \ \ }M_{k}=\left(
\begin{array}
[c]{cc}%
r_{1} & f^{\ast}\\
f & r_{2}%
\end{array}
\right)
\end{equation}

with temperature-dependent parameters:

\begin{align}
r_{1}^{\left(  MSW\right)  }  &  =2\left(  J_{1}\Gamma_{2}+m_{k}DG_{2}\right)
+J_{2}\left\{  \Gamma_{5}\left[  \left(  \alpha+1\right)  -2\lambda\eta
_{k}\right]  +2\eta_{k}\left(  1-\lambda^{2}\right)  \left(  \Gamma_{3}%
+\Gamma_{4}-\Gamma_{1}\right)  \right\}  +\frac{A}{2}\Gamma_{6} \nonumber \\
r_{2}^{\left(  MSW\right)  }  &  =2\left(  J_{1}\Gamma_{3}+m_{k}DG_{2}\right)
+\frac{A}{2}\Gamma_{7}\nonumber\\
h_{x}^{\left(  MSW\right)  }  &  =2\gamma_{k}J_{1}\Gamma_{1}\nonumber\\
h_{y}^{\left(  MSW\right)  }  &  =2\left(  \gamma_{k}J_{1}G_{2}+m_{k}%
D\Gamma_{4}\right) \label{r_renorm}
\end{align}

To obtain temperature-dependent expressions for the mean-field parameters
$\Gamma_{i}$ (or equivalently, $g_{i}$), we Fourier transform the thermal
averages $\left\langle a_{i}^{\dagger}a_{j}\right\rangle $, $\left\langle
b_{i}^{\dagger}b_{j}\right\rangle $, $\left\langle a_{i}b_{j}\right\rangle $
and $\left\langle b_{i}a_{j}\right\rangle ,$ and make a change of basis using

\begin{equation}
\psi_{k}=T_{k}\varphi_{k}%
\end{equation}

with $\psi_{k}^{\dagger}=\left(
\begin{array}
[c]{cccc}%
a_{k}^{\dagger} & b_{-k} & a_{-k} & b_{k}^{\dagger}%
\end{array}
\right)  $ being the original basis, and $\varphi_{k}^{\dagger}=\left(
\begin{array}
[c]{cccc}%
\alpha_{k}^{\dagger} & \beta_{-k} & \alpha_{-k} & \beta_{k}^{\dagger}%
\end{array}
\right)  $ being a new basis. The matrix $T_{k}$ is given by Eq. (\ref{Tk}). As
mentioned before, in a particle-hole Hamiltonian, the Hilbert space is
duplicated, so we can find an irreducible $2\times2$ representation for the
transformation above:

\begin{equation}
\left(
\begin{array}
[c]{c}%
a_{k}\\
b_{-k}^{\dagger}%
\end{array}
\right)  =\left(
\begin{array}
[c]{cc}%
u^{\ast} & -v\\
-v^{\ast} & u
\end{array}
\right)  \left(
\begin{array}
[c]{c}%
\alpha_{k}\\
\beta_{-k}^{\dagger}%
\end{array}
\right)  \label{transform}%
\end{equation}

The thermal averages, after a change of basis, can be written in terms of the
parameters $u$ and $v$ and the occupation number of the bands $n_{k}^{a,\beta
}$. For instance:%

\begin{align}
\left\langle a_{i}^{\dagger}a_{i}\right\rangle  &  =\frac{2}{N}\sum
_{k}\left\langle a_{k}^{\dagger}a_{k}\right\rangle =\frac{2}{N}\sum_{k}\left[
\left\vert u\right\vert ^{2}n_{k}^{\alpha}+\left\vert v\right\vert ^{2}\left(
1+n_{k}^{\beta}\right)  \right] \nonumber\\
\left\langle b_{i}^{\dagger}b_{i}\right\rangle  &  =\frac{2}{N}\sum
_{k}\left\langle b_{k}^{\dagger}b_{k}\right\rangle =\frac{2}{N}\sum_{k}\left[
\left\vert v\right\vert ^{2}\left(  1+n_{k}^{\alpha}\right)  +\left\vert
u\right\vert ^{2}n_{k}^{\beta}\right]
\end{align}

where $N$ is the total number of sites. Here,

\begin{align}
n_{k}^{\alpha}  &  =\left\langle \alpha_{k}^{\dagger}\alpha_{k}\right\rangle
=\left[  \exp\left(  E_{k}^{\beta }/k_{B}T\right)  -1\right]
^{-1}\nonumber\\
n_{k}^{\beta}  &  =\left\langle \beta_{k}^{\dagger}\beta_{k}\right\rangle
=\left[  \exp\left(  E_{k}^{  \alpha  }/k_{B}T\right)  -1\right]
^{-1}%
\end{align}

are the Bose-Einstein distributions. The terms $\left\langle \alpha
_{k}^{\dagger}\alpha_{k}\right\rangle $ and $\left\langle \beta_{k}^{\dagger
}\beta_{k}\right\rangle $ are the only thermal averages in the new basis that
are not zero. To illustrate that, let's consider a term of the form%

\begin{equation}
\left\langle \alpha_{k}\beta_{-k}\right\rangle =\frac{1}{Z}\sum_{n}e^{-\beta
E_{k}^{  n  }}\left\langle n\right\vert \alpha_{k}\beta
_{-k}\left\vert n\right\rangle
\end{equation}

The matrix element $\left\langle n\right\vert \alpha_{k}\beta_{-k}\left\vert
n\right\rangle $ is an overlap of the state $\beta_{-k}\left\vert
n\right\rangle $ and $\alpha_{k}^{\dagger}\left\vert n\right\rangle $. Both
states are eigenstates of $H_{k}$, but as they do not have identical sets of
occupation numbers of $\alpha$ and $\beta$ bosons, their overlap is zero.
Hence $\left\langle \alpha_{k}\beta_{-k}\right\rangle $ and its complex
conjugate are zero. This occurs for every other thermal average, except
$\left\langle \alpha_{k}^{\dagger}\alpha_{k}\right\rangle $ and $\left\langle
\beta_{k}^{\dagger}\beta_{k}\right\rangle $.

Performing the procedure detailed above, we obtain temperature-dependent
expressions for the $g_{i}$ and $G_{i}$ parameters:

\begin{align}
g_{1}  &  =\frac{2}{N}\sum_{k}\left[  \left\vert u\right\vert ^{2}%
n_{k}^{\alpha}+\left\vert v\right\vert ^{2}\left(  1+n_{k}^{\beta}\right)
\right] \nonumber\\
g_{2}  &  =\frac{2}{N}\sum_{k}\left[  \left\vert u\right\vert ^{2}n_{k}%
^{\beta}+\left\vert v\right\vert ^{2}\left(  1+n_{k}^{\alpha}\right)  \right]
\nonumber\\
g_{5}  &  =\frac{2}{N}\sum_{k}\eta_{k}\left[  \left\vert u\right\vert
^{2}n_{k}^{\alpha}+\left\vert v\right\vert ^{2}\left(  1+n_{k}^{\beta}\right)
\right] \nonumber\\
g_{6}  &  =\frac{2}{N}\sum_{k}\eta_{k}\left[  \left\vert u\right\vert
^{2}\left(  1+n_{k}^{\alpha}\right)  +\left\vert v\right\vert ^{2}n_{k}%
^{\beta}\right] \nonumber\\
G_{1}  &  =-\frac{2}{N}\sum_{k}\gamma_{k}\,xv\left(  1+n_{k}^{\alpha}%
+n_{k}^{\beta}\right) \nonumber\\
G_{2}  &  =\,\frac{2}{N}\sum_{k}\gamma_{k}\,yv\left(  1+n_{k}^{\alpha}%
+n_{k}^{\beta}\right)  \label{gs}%
\end{align}

where we defined $u=x+iy$. The sublattice (staggered) magnetization is given by%

\begin{equation}
m=S-\left\langle a_{i}^{\dagger}a_{i}\right\rangle =S-\frac{2}{N}\sum
_{k}\left[  \left\vert u\right\vert ^{2}n_{k}^{\alpha}+\left\vert v\right\vert
^{2}\left(  1+n_{k}^{\beta}\right)  \right]
\end{equation}

The temperature dependence comes from the Bose-Einstein factors. In the
continuum limit, the summation becomes an integral over the Brillouin zone:

\begin{equation}
\frac{2}{N}\sum_{\mathbf{k}}\left[  \clubsuit\right]  \rightarrow
\int\limits_{BZ}\frac{d^{2}k}{\left(  2\pi\right)  ^{2}}\left[  \clubsuit
\right]
\end{equation}

For each temperature, we can obtain the $\Gamma_{i}$ factors self-consistently from the Eqs. (\ref{gs}) and Eqs. (\ref{Gamma}). These terms renormalize the Hamiltonian.

In summary, the effective Hamiltonian becomes temperature-dependent when we
include quartic terms through a mean-field decoupling. For each temperature,
coefficients $\Gamma_{i}$ that renormalize the Hamiltonian parameters can be
obtained self-consistently through (\ref{Gamma}) and (\ref{gs}). All the
equations in Section 5 remain the same, but with the renormalized
parameters $r_{1},$ $r_{2}$ and $f$, following Eqs. (\ref{r_renorm}).

\section{Berry curvature of a Bogoliubov-de Gennes Hamiltonian}
\renewcommand{\theequation}{\thesection{.}\arabic{equation}}
\setcounter{equation}{0}

We start with the Berry curvature given by Eq. (\ref{berry_c}) \cite{Samajdar2019}:%

\begin{equation}
\Omega_{xy}^{n}\left(  \mathbf{k}\right)  =i\sum_{\mu\nu}\varepsilon_{\mu\nu
}\left[  \eta\frac{\partial T_{k}^{\dagger}}{\partial k_{\mu}}\eta
\frac{\partial T_{k}}{\partial k_{\nu}}\right]  _{nn}%
\end{equation}

The matrices $\eta$ and $T_{k}$ are block diagonal:%

\begin{equation}
\eta=\left(
\begin{array}
[c]{cccc}%
1 & 0 & 0 & 0\\
0 & -1 & 0 & 0\\
0 & 0 & -1 & 0\\
0 & 0 & 0 & 1
\end{array}
\right)  =\left(
\begin{array}
[c]{cc}%
\sigma_{z} & 0\\
0 & -\sigma_{z}%
\end{array}
\right)
\end{equation}

{\color{black}
\begin{equation}
T_{k}=\left(
\begin{array}
[c]{cccc}%
u_k^{\ast} & -v_k & 0 & 0\\
-v_k^{\ast} & u_k & 0 & 0\\
0 & 0 & u_{-k} & -v_{-k}^{\ast}\\
0 & 0 & -v_{-k} & u_{-k}^{\ast}%
\end{array}
\right)  =\left(
\begin{array}
[c]{cc}%
T_{\alpha} & 0\\
0 & T_{\beta}%
\end{array}
\right)  , \label{block_Tk}%
\end{equation}
}

so we can perform a block multiplication:%

\begin{equation}
\eta\frac{\partial T_{k}^{\dagger}}{\partial k_{\mu}}\eta\frac{\partial T_{k}%
}{\partial k_{\nu}}=\left(
\begin{array}
[c]{cc}%
\sigma_{z}\frac{\partial T_{\alpha}^{\dagger}}{\partial k_{\mu}}\sigma
_{z}\frac{\partial T_{\alpha}}{\partial k_{\nu}} & 0\\
0 & \sigma_{z}\frac{\partial T_{\beta}^{\dagger}}{\partial k_{\mu}}\sigma
_{z}\frac{\partial T_{\beta}}{\partial k_{\nu}}%
\end{array}
\right)
\end{equation}

The expression for the Berry curvature can be reduced to 2$\times$2 representation%

\begin{equation}
\Omega_{xy}^{\left(  n\right)  }\left(  \mathbf{k}\right)  =i\sum_{\mu\nu
}\varepsilon_{\mu\nu}\left(  \sigma_{z}\frac{\partial T_{\alpha,\beta
}^{\dagger}}{\partial k_{\mu}}\sigma_{z}\frac{\partial T_{\alpha,\beta}%
}{\partial k_{\nu}}\right)  _{nn}%
\end{equation}

where we can choose the $\alpha$ or $\beta$ sector. We focus on the particle
states, remembering that for the $\alpha$-sector, it corresponds to the first
column of $T_{\alpha}$ ($n=1$), and for the $\beta$-sector, to the second
column of $T_{\beta}$ ($n=2$).

Focusing on the $\alpha$-sector first, we have:%

\begin{equation}
\Omega_{xy}^{\alpha}\left(  \mathbf{k}\right)  =i\sum_{\mu\nu}\varepsilon_{\mu
\nu}\left(  \sigma_{z}\frac{\partial T_{\alpha}^{\dagger}}{\partial k_{\mu}%
}\sigma_{z}\frac{\partial T_{\alpha}}{\partial k_{\nu}}\right)  _{11}%
\end{equation}

Using $T_{\alpha}=\left(
\begin{array}
[c]{cc}%
u_k^{\ast} & -v_k\\
-v_k^{\ast} & u_k
\end{array}
\right)  $ (see Section 5), we have \textcolor{black}{(we suppress the index $k$ for clearer notation)}:

\begin{align}
\sigma_{z}\frac{\partial T_{\alpha}^{\dagger}}{\partial k_{x}}\sigma_{z}%
\frac{\partial T_{\alpha}}{\partial k_{y}}  &  =\left(
\begin{array}
[c]{cc}%
\frac{\partial u}{\partial k_{x}} & -\frac{\partial v}{\partial k_{x}%
}\\
\frac{\partial v^{\ast}}{\partial k_{x}} & -\frac{\partial u^{\ast}}{\partial k_{x}}%
\end{array}
\right)  \left(
\begin{array}
[c]{cc}%
\frac{\partial u^{\ast}}{\partial k_{y}} & -\frac{\partial v}{\partial k_{y}%
}\\
\frac{\partial v^{\ast}}{\partial k_{y}} & -\frac{\partial u}{\partial k_{y}}%
\end{array}
\right) \nonumber\\
\rightarrow\left[  \sigma_{z}\frac{\partial T_{\alpha}^{\dagger}}{\partial
k_{x}}\sigma_{z}\frac{\partial T_{\alpha}}{\partial k_{y}}\right]  _{11}  &
=\frac{\partial u}{\partial k_{x}}\frac{\partial u^{\ast}}{\partial k_{y}%
}-\frac{\partial v}{\partial k_{x}}\frac{\partial v^{\ast}}{\partial
k_{y}}%
\end{align}

and we can write

\begin{align}
\sum_{\mu\nu}\varepsilon_{\mu\nu}\left(  \sigma_{z}\frac{\partial T_{\alpha
}^{\dagger}}{\partial k_{\mu}}\sigma_{z}\frac{\partial T_{\alpha}}{\partial
k_{\nu}}\right)  _{11}&=\left(  \frac{\partial u}{\partial k_{x}}\frac{\partial
u^{\ast}}{\partial k_{y}}-\frac{\partial v}{\partial k_{x}}%
\frac{\partial v^{\ast}}{\partial k_{y}}\right)  -\left(  \frac{\partial
u}{\partial k_{y}}\frac{\partial u^{\ast}}{\partial k_{x}}-\frac{\partial
v}{\partial k_{y}}\frac{\partial v^{\ast}}{\partial k_{x}}\right) \nonumber \\
&=\left(  \frac{\partial u}{\partial k_{x}}\frac{\partial
u^{\ast}}{\partial k_{y}}-\frac{\partial v}{\partial k_{x}}%
\frac{\partial v^{\ast}}{\partial k_{y}}\right) - C.C.
\end{align}

The expression for the Berry curvature of the particle $\alpha$ state reduces to

\begin{align}
\Omega_{xy}^{\alpha}\left(  \mathbf{k}\right)   &  =i\left[ 
\left(  \frac{\partial u}{\partial k_{x}}\frac{\partial
u^{\ast}}{\partial k_{y}}-\frac{\partial v}{\partial k_{x}} \frac{\partial v^{\ast}}{\partial k_{y}}\right) - C.C. \right] = i \left[ 2i \operatorname{Im} \left(  \frac{\partial u}{\partial k_{x}}\frac{\partial
u^{\ast}}{\partial k_{y}}-\frac{\partial v}{\partial k_{x}} \frac{\partial v^{\ast}}{\partial k_{y}}\right)  \right] \nonumber\\
\Omega_{xy}^{\alpha}\left(  \mathbf{k}\right)   &  =-2\operatorname{Im}\left(  \frac{\partial u}{\partial k_{x}}\frac{\partial
u^{\ast}}{\partial k_{y}}-\frac{\partial v}{\partial k_{x}} \frac{\partial v^{\ast}}{\partial k_{y}}\right) 
\end{align}

Using $u=e^{i\phi}\cosh\left(  \frac{\theta}{2}\right)$ and $v=\left( \frac{r-w}{2w}\right)^{1/2}$ we get (Im $\frac{\partial v}{\partial k_{x}} \frac{\partial v^{\ast}}{\partial k_{y}}=0$ because $v$ is real):

\begin{align}
\frac{\partial u}{\partial k_{x}}  &  =e^{i\phi}\left[  i\cosh\frac{\theta}%
{2}\left(  \frac{\partial\phi}{\partial k_{x}}\right)  +\frac{1}{2}\sinh
\frac{\theta}{2}\left(  \frac{\partial\theta}{\partial k_{x}}\right)  \right] \nonumber \\
\frac{\partial u^{\ast}}{\partial k_{y}}  &  =e^{-i\phi}\left[  -i\cosh
\frac{\theta}{2}\left(  \frac{\partial\phi}{\partial k_{y}}\right)  +\frac
{1}{2}\sinh\frac{\theta}{2}\left(  \frac{\partial\theta}{\partial k_{y}%
}\right)  \right]
\end{align}

and using $\sinh\left(  \frac{\theta}{2}\right)  \cosh\left(  \frac{\theta}%
{2}\right)  =\frac{1}{2}\sinh\theta$ we finally get%

{\color{black}
\begin{equation}
\Omega_{xy}^{\alpha}\left(  \mathbf{k}\right)  =-\frac{1}{2}\sinh\theta_{k}\left(
\frac{\partial\phi_{k}}{\partial k_{x}}\frac{\partial\theta_{k}}{\partial k_{y}}%
-\frac{\partial\phi_{k}}{\partial k_{y}}\frac{\partial\theta_{k}}{\partial k_{x}%
}\right) \label{berry_1}
\end{equation}
}

For the $\beta$-sector, the initial expression is

\begin{equation}
\Omega_{xy}^{\beta}\left(  \mathbf{k}\right)  =i\sum_{\mu\nu}\varepsilon_{\mu\nu
}\left(  \sigma_{z}\frac{\partial T_{\beta}^{\dagger}}{\partial k_{\mu}}%
\sigma_{z}\frac{\partial T_{\beta}}{\partial k_{\nu}}\right)  _{22}%
\end{equation}

{\color{black}
And noting that $T_{\beta}(\mathbf{k})=T_{\alpha}^{\ast}(\mathbf{-k})$, it is easy to show that

\begin{equation}
\Omega_{xy}^{\beta}\left(  \mathbf{k}\right)  = -\frac{1}{2}\sinh\theta_{-k}\left(  \frac{\partial\phi_{-k}}{\partial
k_{x}}\frac{\partial\theta_{-k}}{\partial k_{y}}-\frac{\partial\phi_{-k}}{\partial
k_{y}}\frac{\partial\theta_{-k}}{\partial k_{x}}\right) \label{berry_2}
\end{equation}

The general relation between the Berry curvatures of the two bands is $\Omega_{xy}^{\beta}\left(  \mathbf{k}\right)=\Omega_{xy}^{\alpha}\left(  \mathbf{-k}\right)$. When these are even functions ($\Omega_{xy}^{\alpha}\left(  \mathbf{k}\right)=\Omega_{xy}^{\alpha}\left(  \mathbf{-k}\right)$, as it is the case of the Union Jack lattice), both Berry curvatures have the same sign: $\Omega_{xy}^{\beta}\left(  \mathbf{k}\right)=\Omega_{xy}^{\alpha}\left(  \mathbf{k}\right)$. But when they are odd functions ($\Omega_{xy}^{\alpha}\left(  \mathbf{k}\right)=-\Omega_{xy}^{\alpha}\left(  \mathbf{-k}\right)$, as it is the case of the brick-wall lattice) the Berry curvatures have opposite signs: $\Omega_{xy}^{\beta}\left(  \mathbf{k}\right)=-\Omega_{xy}^{\alpha}\left(  \mathbf{k}\right)$

It is possible to show that the hole-states, which correspond to the second
and third columns of $T_{k}$, have opposite Berry curvature in the same band index as a consequence of particle-hole symmetry:

\begin{align}
\Omega_{xy}^{\alpha\left( hole\right)  }\left(  \mathbf{k}\right)  &=-\Omega
_{xy}^{\alpha \left( particle\right)  }\left(  \mathbf{k}\right) \nonumber \\
\Omega_{xy}^{\beta\left( hole\right)  }\left(  \mathbf{k}\right)  &=-\Omega
_{xy}^{\beta \left( particle\right)  }\left(  \mathbf{k}\right) 
\end{align}
}

Evaluating Eq. (\ref{berry_1}) with the definitions in Section 4, we arrive at the expression:

\begin{align}
\Omega^\alpha_{xy}(\mathbf{k})=-\frac{1}{2}\frac{1}{w^{3}\left\vert f\right\vert }  &\left\{
\left(  h_{x}\frac{\partial h_{y}}{\partial k_{x}}-h_{y}\frac{\partial h_{x}%
}{\partial k_{x}}\right)  \left[  \frac{r}{\left\vert f\right\vert }\left(
h_{x}\frac{\partial h_{x}}{\partial k_{y}}+h_{y}\frac{\partial h_{y}}{\partial
k_{y}}\right)  -\left\vert f\right\vert \frac{\partial r}{\partial k_{y}%
}\right] \right. \nonumber \\ 
&- \left. \left(  h_{x}\frac{\partial h_{y}}{\partial k_{y}}-h_{y}\frac{\partial h_{x}%
}{\partial k_{y}}\right)  \left[  \frac{r}{\left\vert f\right\vert }\left(
h_{x}\frac{\partial h_{x}}{\partial k_{x}}+h_{y}\frac{\partial h_{y}}{\partial
k_{x}}\right)  -\left\vert f\right\vert \frac{\partial r}{\partial k_{x}%
}\right] \right\} \label{BerryCurvature}
\end{align}

\textcolor{black}{and from this we can plot the Berry curvature of any system, knowing the Hamiltonian parameters $h_x$, $h_y$ and $r$.} \textcolor{black}{We stress here that the imaginary part of the Hamiltonian comes from $h_y$. From the expression above, we see it is crucial that $h_y \neq 0 $ for a non-null Berry curvature. In other words: an imaginary term in the Hamiltonian is necessary for the system to have a non-null Berry curvature.}

\end{appendices}

\newpage

\bibliographystyle{ieeetr}
\bibliography{refs_UJ_reordered}

\end{document}